\documentclass[ALICE,manyauthors]{cernphprep}
\usepackage[comma,square,numbers,sort&compress,merge]{natbib}
\usepackage{hyperref}
\usepackage{lineno}
\usepackage{xspace}
\usepackage{stackengine}
\usepackage[T1]{fontenc}

\newcommand\xrowht[2][0]{\addstackgap[.5\dimexpr#2\relax]{\vphantom{#1}}}

\begin{document}
%

\newcommand{\pp}           {pp\xspace}
\newcommand{\ppbar}        {\mbox{$\mathrm {p\overline{p}}$}\xspace}
\newcommand{\XeXe}         {\mbox{Xe--Xe}\xspace}
\newcommand{\PbPb}         {\mbox{Pb--Pb}\xspace}
\newcommand{\pA}           {\mbox{pA}\xspace}
\newcommand{\pPb}          {\mbox{p--Pb}\xspace}
\newcommand{\AuAu}         {\mbox{Au--Au}\xspace}
\newcommand{\dAu}          {\mbox{d--Au}\xspace}

\newcommand{\s}            {\ensuremath{\sqrt{s}}\xspace}
\newcommand{\snn}          {\ensuremath{\sqrt{s_{\mathrm{NN}}}}\xspace}
\newcommand{\pt}           {\ensuremath{p_{\rm T}}\xspace}
\newcommand{\meanpt}       {$\langle p_{\mathrm{T}}\rangle$\xspace}
\newcommand{\ycms}         {\ensuremath{y_{\rm CMS}}\xspace}
\newcommand{\ylab}         {\ensuremath{y_{\rm lab}}\xspace}
\newcommand{\etarange}[1]  {\mbox{$\left | \eta \right |~<~#1$}}
\newcommand{\yrange}[1]    {\mbox{$\left | y \right |~<~#1$}}
\newcommand{\dndy}         {\ensuremath{\mathrm{d}N_\mathrm{ch}/\mathrm{d}y}\xspace}
\newcommand{\dndeta}       {\ensuremath{\mathrm{d}N_\mathrm{ch}/\mathrm{d}\eta}\xspace}
\newcommand{\avdndeta}     {\ensuremath{\langle\dndeta\rangle}\xspace}
\newcommand{\dNdy}         {\ensuremath{\mathrm{d}N_\mathrm{ch}/\mathrm{d}y}\xspace}
\newcommand{\Npart}        {\ensuremath{N_\mathrm{part}}\xspace}
\newcommand{\Ncoll}        {\ensuremath{N_\mathrm{coll}}\xspace}
\newcommand{\dEdx}         {\ensuremath{\textrm{d}E/\textrm{d}x}\xspace}
\newcommand{\RpPb}         {\ensuremath{R_{\rm pPb}}\xspace}

\newcommand{\nineH}        {$\sqrt{s}~=~0.9$~Te\kern-.1emV\xspace}
\newcommand{\seven}        {$\sqrt{s}~=~7$~Te\kern-.1emV\xspace}
\newcommand{\twoH}         {$\sqrt{s}~=~0.2$~Te\kern-.1emV\xspace}
\newcommand{\twosevensix}  {$\sqrt{s}~=~2.76$~Te\kern-.1emV\xspace}
\newcommand{\five}         {$\sqrt{s}~=~5.02$~Te\kern-.1emV\xspace}
\newcommand{\twosevensixnn}{$\sqrt{s_{\mathrm{NN}}}~=~2.76$~Te\kern-.1emV\xspace}
\newcommand{\fivenn}       {$\sqrt{s_{\mathrm{NN}}}~=~5.02$~Te\kern-.1emV\xspace}
\newcommand{\LT}           {L{\'e}vy-Tsallis\xspace}
\newcommand{\GeVc}         {Ge\kern-.1emV/$c$\xspace}
\newcommand{\MeVc}         {Me\kern-.1emV/$c$\xspace}
\newcommand{\TeV}          {Te\kern-.1emV\xspace}
\newcommand{\GeV}          {Ge\kern-.1emV\xspace}
\newcommand{\MeV}          {Me\kern-.1emV\xspace}
\newcommand{\GeVmass}      {Ge\kern-.2emV/$c^2$\xspace}
\newcommand{\MeVmass}      {Me\kern-.2emV/$c^2$\xspace}
\newcommand{\lumi}         {\ensuremath{\mathcal{L}}\xspace}

\newcommand{\ITS}          {\rm{ITS}\xspace}
\newcommand{\TOF}          {\rm{TOF}\xspace}
\newcommand{\ZDC}          {\rm{ZDC}\xspace}
\newcommand{\ZDCs}         {\rm{ZDCs}\xspace}
\newcommand{\ZNA}          {\rm{ZNA}\xspace}
\newcommand{\ZNC}          {\rm{ZNC}\xspace}
\newcommand{\SPD}          {\rm{SPD}\xspace}
\newcommand{\SDD}          {\rm{SDD}\xspace}
\newcommand{\SSD}          {\rm{SSD}\xspace}
\newcommand{\TPC}          {\rm{TPC}\xspace}
\newcommand{\TRD}          {\rm{TRD}\xspace}
\newcommand{\VZERO}        {\rm{V0}\xspace}
\newcommand{\VZEROA}       {\rm{V0A}\xspace}
\newcommand{\VZEROC}       {\rm{V0C}\xspace}
\newcommand{\Vdecay} 	   {\ensuremath{V^{0}}\xspace}

\newcommand{\ee}           {\ensuremath{e^{+}e^{-}}} 
\newcommand{\pip}          {\ensuremath{\pi^{+}}\xspace}
\newcommand{\pim}          {\ensuremath{\pi^{-}}\xspace}
\newcommand{\kap}          {\ensuremath{\rm{K}^{+}}\xspace}
\newcommand{\kam}          {\ensuremath{\rm{K}^{-}}\xspace}
\newcommand{\pbar}         {\ensuremath{\rm\overline{p}}\xspace}
\newcommand{\kzero}        {\ensuremath{{\rm K}^{0}_{\rm{S}}}\xspace}
\newcommand{\lmb}          {\ensuremath{\Lambda}\xspace}
\newcommand{\almb}         {\ensuremath{\overline{\Lambda}}\xspace}
\newcommand{\Om}           {\ensuremath{\Omega^-}\xspace}
\newcommand{\Mo}           {\ensuremath{\overline{\Omega}^+}\xspace}
\newcommand{\X}            {\ensuremath{\Xi^-}\xspace}
\newcommand{\Ix}           {\ensuremath{\overline{\Xi}^+}\xspace}
\newcommand{\Xis}          {\ensuremath{\Xi^{\pm}}\xspace}
\newcommand{\Oms}          {\ensuremath{\Omega^{\pm}}\xspace}
\newcommand{\degree}       {\ensuremath{^{\rm o}}\xspace}


\begin{titlepage}
\PHyear{2021}       
\PHnumber{144}      
\PHdate{18 July}  

\title{Study of very forward energy and its correlation with particle production at midrapidity in pp and p--Pb collisions at the LHC}
\ShortTitle{Very forward energy in pp and p--Pb collisions}   

\Collaboration{ALICE Collaboration\thanks{See Appendix~\ref{app:collab} for the list of collaboration members}}
\ShortAuthor{ALICE Collaboration} 

\begin{abstract}

The energy deposited at very forward rapidities (very forward energy)
is a powerful tool for cha\-rac\-te\-ri\-sing proton fragmentation in pp and p--Pb collisions. The correlation of very forward energy with particle production at midrapidity provides direct insights into the initial stages and the subsequent evolution of the collision.
Furthermore, the correlation with the production of particles with large transverse momenta at midrapidity provides information complementary to the measurements of the underlying event, which are usually interpreted in the framework of models implementing centrality-dependent multiple parton interactions.

Results about very forward energy, measured by the ALICE zero degree calorimeters (ZDCs), and its dependence on the activity measured at midrapidity in pp collisions at $\sqrt{s}=13$~TeV and in p--Pb collisions at $\sqrt{s_{\rm{NN}}}=8.16$~TeV are discussed.
The measurements performed in pp collisions are compared with the expectations of three hadronic interaction event generators: PYTHIA~6 (Perugia 2011 tune), PYTHIA~8 (Monash tune), and EPOS~LHC.
These results provide new constraints on the validity of models in describing the beam remnants at very forward rapidities, where perturbative QCD cannot be used.
\end{abstract}
\end{titlepage}

\setcounter{page}{2} 


\section{Introduction}

During the last decade, the study of bulk properties of proton--proton (pp) and proton--nucleus (pA) collisions at LHC energies has attracted increasing interest. Effects expected to occur only in heavy-ion collisions, such as collective fluid-like behaviour~\cite{ppridge, pAridge} and strangeness enhancement~\cite{sNature}, are already seen in pp collisions. In particular, strangeness enhancement already occurs at small multiplicity values.
The strength of these effects increases steadily with the final state multiplicity going from pp to p--Pb up to peripheral Pb--Pb collisions. As a consequence, minimum-bias (MB) pp collisions are currently being studied, focusing on the dependence of experimental observables on the charged-particle multiplicity measured at midrapidity.
In phenomenological models inspired by quantum chromodynamics (QCD),
high-multiplicity pp and pA events are generated in collisions with smaller than average impact parameter, $b$ (the distance between the centres of the two colliding hadrons). In these central collisions, the number of partonic interactions and, hence, the probability for partonic scatterings with large momentum transfer, is enhanced. On the other hand, by requesting the production of a particle with large transverse momentum,
events with larger than average multiplicity
are selected~\cite{ALICEmultptsel}.
In order to constrain the initial state of pp and p--Pb collisions, the correlation between observables measured in rapidity regions that are causally disconnected in the evolution of the system following the collision are studied. In particular, this is the case for the very forward (``zero degree'') energy and central rapidity particle production.
The study of these correlations also addresses the fundamental question of how the energy of the colliding protons is transferred from beam to central rapidities. In models where the initial state is described by the impact parameter or initial matter overlap, an increased energy transfer results naturally from the increased number of parton--parton scatterings.

The processes involved in forward hadron production at high energies are crucial for simulations of high-energy cosmic-ray interactions~\cite{cosmics}. Predictions from different models describing hadronic interactions suffer from large uncertainties. Experimental results on the production of forward baryons have therefore a key role in the simulation of cosmic-ray showers. Measurements at colliders can provide unique information for the tuning of the models. In particular, baryon production and remnant break-up have been addressed as possible mechanism to explain the observed muon production at ground level~\cite{Pierog:2006qv, Drescher:2007hc}.

At LHC energies, forward neutron production has been measured by the LHCf collaboration~\cite{LHCf:2015nel, LHCf:2020hjf, LHCf:2018gbv}.
In ALICE, the zero degree calorimeters (ZDCs) measure the energy emitted at rapidities close to that of the beam, covering the beam fragmentation regions.  The very forward energy measured by the ZDCs can be correlated with the charged-particle multiplicity and the probability of producing a particle with a large transverse momentum, \pt, at midrapidity. Measurements of very forward energy as a function of the \pt of the leading particle (the charged particle with the highest transverse momentum) at midrapidity complement the studies of the underlying event (UE). In the first case, particle production is related to an observable separated in pseudorapidity, while in the UE measurement particle production is measured in a region separated in azimuthal angle (``transverse region").

This article presents the first measurements of the dependence of the very forward energy on midrapidity particle production in pp collisions at a centre-of-mass energy $\sqrt{s}=13$~TeV and in p--Pb collisions at a centre-of-mass-energy per nucleon pair $\sqrt{s_{\rm{NN}}}=8.16$~TeV.
The article is organised as follows: in Sec.~\ref{sec.2} the main ALICE subsystems used for this analysis are concisely described. The data sample, the event and track selection criteria are discussed in Sec.~\ref{sec.3}. Section~\ref{sec.4} presents results about the production of energy at forward and backward rapidities in pp collisions. In Sec.~\ref{sec.5} results on the correlation between very forward energy and particle production at midrapidity in pp and in p--Pb collisions are presented and discussed, together with results on the forward leading-baryon production in connection with the UE measurements in pp collisions.

\section{ALICE detectors}\label{sec.2}

A detailed description of the ALICE detector and its performance can be found in Refs.~\cite{ALICEdet, ALICEperf}.
The sub-detectors used for the present analysis are the inner tracking system (ITS), the time projection chamber (TPC) and the V0 detectors, all located inside a 0.5~T uniform magnetic field, and the ZDCs.

The silicon pixel detector (SPD)~\cite{ALICEdet} makes up the two cylindrical innermost layers of the ITS and consists of hybrid silicon pixel assemblies covering the pseudorapidity range $|\eta|<2$ for the inner layer and $|\eta|<1.4$ for the outer layer for collisions occurring at the nominal interaction point (IP). The SPD is used to measure the charged-particle multiplicity at midrapidity using
tracklets, short track segments formed using information on the position of the primary vertex and two hits, one on each of the SPD layers.
To exploit the full particle tracking, the four external layers of the ITS, composed by two layers of silicon drift detectors (SDD) and two layers of double-sided silicon micro-strip detectors (SSD), were also used.
The TPC~\cite{TPC} is the main tracking detector and it covers a pseudorapidity range of about $|\eta|<0.9$. In order to avoid border effects, in this analysis the fiducial pseudorapidity region has been restricted to $|\eta|<0.8$.
Charged-particle tracks are formed by combining the hits in the ITS and the reconstructed clusters in the TPC.
The trigger signal is provided by the V0~\cite{V0} counters, two arrays of 32 scintillator tiles each, covering the full azimuth within $-3.7<\eta<-1.7$ for V0C,  located on the same side as for the ALICE dimuon arm (C-side), and $2.8<\eta<5.1$ for V0A, placed on the opposite side (A-side).
An alternative trigger condition can be provided by the ALICE diffractive (AD) detector~\cite{AD}, a hodoscope of plastic scintillators covering the pseudorapidity ranges $-7.0<\eta<-4.9$ and $4.8<\eta<6.3$.

The main detectors used in this analysis are the ZDCs. Two identical systems, each comprising a neutron (ZN) and a proton (ZP) calorimeter of the ``spaghetti'' type, are placed at $\pm$112.5~m from the ALICE IP, on both sides.
In nucleus--nucleus collisions, the ZDCs detect the energy carried by the non-interacting (spectator) nucleons and they are used to estimate the centrality of the collision~\cite{ALICEcentrAA}.
In proton--nucleus collisions the ZDCs provide an unbiased centrality selection~\cite{ALICEcentrpA}.
In proton--proton collisions, the ZDCs are usually switched off, not only to prevent their aging but also because the typical beam conditions in pp data taking are characterised by large crossing angles that drastically affect the ZDC acceptance. However, the acceptance of ZN is not affected provided that the vertical half-crossing angle is smaller than $+60 \ \mu$rad for a nominal vertex vertical position on the LHC axis ($y_{\rm{vtx}} = 0$~mm).

In pp collisions, the energy detected by ZN is mainly due to neutrons with a contribution from photons at low energies, while the ZP signal is essentially due to protons with a smaller contribution due to positive pions.
The ZN calorimeters cover the pseudorapidity range $|\eta|>8.8$ and within this range they have a flat efficiency. For the ZP calorimeters the purely geometric coverage, $6.5<|\eta|<7.4$, is not significant since the actual pseudorapidity interval covered by ZP depends strongly on the LHC beam optics settings. The covered pseudorapidity range has been studied through fast Monte Carlo (MC) simulations taking into account the magnetic field settings and the geometrical apertures of the LHC beam pipe element for pp collisions at $\sqrt{s}=13$~TeV and found to be  $7.8<|\eta|<12.9$.
In Pb--Pb collisions and in p--Pb collisions in the Pb-fragmentation region, the energy calibrations of ZN and ZP spectra are performed using the narrow peaks from the detection of single nucleons. Contrarily, in pp collisions and in p--Pb collisions in the p-fragmentation region, there are no peaks in the spectra and there is no reliable way to calibrate the spectra in energy units without introducing model dependencies and large uncertainties. The use of self-normalised quantities, namely signals normalised to their average minimum-bias value, allows one to overcome this issue, since they coincide with the self-normalised energy ratios and are therefore directly comparable to model predictions.

The ZDCs are equipped with time-to-digital converters (TDCs) that register the arrival time of particles depositing energy in the detectors, allowing the rejection of events without signal in the calorimeters (noise).
The event selection fractions for both ZN and ZP are defined as the ratios between the number of events with a signal in the corresponding TDC and the number of MB events triggered by ALICE. The event selection fractions values calculated for the two colliding systems are given in Table~\ref{tab:2}.
\begin{table}[t]
    \caption{ZN and ZP  event selection fractions (see text for details) in pp and p--Pb collisions. For the p--Pb colliding system, both the Pb-fragmentation (Pb) and the p-fragmentation (p) sides are reported.}
    \centering
    \begin{tabular}{|c|c|c|c|}
         \hline
         \xrowht[()]{10pt}
         & pp $\sqrt{s}=$~13 TeV & p--Pb $\sqrt{s_{\rm{NN}}}=8.16$~TeV\\
         \hline
         ZN & 61\% & 96\% (Pb) 43\% (p)\\ \hline
         ZP & 23\% & 82\% (Pb) 15\% (p) \\
         \hline
    \end{tabular}
    \label{tab:2}
\end{table}

\section{Data samples, event selection and models}\label{sec.3}

During the 2015 pp data taking at $\sqrt{s}=13$~TeV, the ZDCs were switched on when a limited half-crossing angle of $+45~\mu$rad in the vertical plane was applied. As discussed in the previous section, this configuration guaranteed that all the neutrons emitted at very forward rapidities were within the ZN geometric acceptance.
Data from p--Pb collisions, collected in 2016 at centre-of-mass energy $\sqrt{s_{\rm{NN}}}=8.16$~TeV, were used to study the forward energy on the p-fragmentation side in pA collisions. The proton beam energy was the same as in pp collisions, therefore comparing the very forward energy in the p-fragmentation side in p--Pb collision to the one detected in pp collisions provides useful information on the proton breakup.

In both collision systems, a MB trigger condition, requiring at least one hit in both V0 detectors, was applied. The V0 timing information was used offline to reject beam--gas interactions.
Events with more than one reconstructed primary interaction vertex (pile-up) were rejected.
The selected events were required to have a reconstructed collision vertex with a position along the beam axis $|z_{\rm {vtx}}|<10$~cm to ensure a uniform performance for midrapidity detectors. After applying these selections, about 4$\times10^{7}$ MB events in pp collisions and 8.3$\times10^{7}$ MB events in p--Pb collisions were retained.

In the analysis of pp collisions, to study the transverse momentum of the leading particle, tracks were reconstructed in $|\eta|<0.8$ (to guarantee that all tracks have the maximal length) and more stringent selection criteria were applied.
In order to ensure good track momentum resolution, the reconstructed tracks were required to have at least 70 reconstructed points (out of a maximum of 159) in the TPC, and two hits in the ITS, with at least one in the SPD. Finally, the $\chi^{2}$ per TPC reconstructed point was required to be less than 4, and tracks originating from kink topologies of weak decays were rejected. These conditions selected tracks with a transverse momentum \pt$>0.15$~GeV/$c$.

The very forward energy was studied as a function of the charged-particle multiplicity measured at midrapidity in two units of pseudorapidity. A primary particle is a particle with a mean proper lifetime $\tau$ larger than 1~cm/$c$, which is either produced directly in the interaction, or from decays of particles with $\tau$ smaller than 1~cm/$c$, excluding particles produced in interactions with material~\cite{ALICE-PUBLIC-2017-005}.
In pp collisions the interval $|\eta|<1$, centered around midrapidity, was considered.  In p--Pb collisions, the nucleon–nucleon centre-of-mass system moves with respect to the ALICE laboratory system with a rapidity of $-0.465$ in the direction of the proton beam. Therefore the pseudorapidity interval $-1.465<\eta<0.535$, covering two units of pseudorapidity around midrapidity in the centre-of-mass system, is used.
The multiplicity selection is based on the number of tracklets measured in the SPD in both collision systems, as described in Refs.~\cite{ALICEppmult, ALICEpAmult}. The average charged-particle multiplicity values, \avdndeta, are listed in Table~\ref{tab:1} for each selected multiplicity interval.
\begin{table}[t]
    \caption{Charged-particle multiplicity classes based on the SPD tracklet estimator and corresponding \avdndeta values in 2 units of pseudorapidity around midrapidity, $|\eta|<1$ in pp collisions at 13~TeV~\cite{ALICEppmult} and $-1.465<\eta<0.535$ in p--Pb collisions at 8.16~TeV~\cite{ALICEpAmult}). The MB values for the two colliding systems are given in the bottom rows.}
    \centering
\begin{minipage}[t]{0.48\linewidth}\centering
    \begin{tabular}{|c|c|}
    \hline
    Multiplicity  & pp 13 TeV \\
    \xrowht[()]{6pt}
    class & \ensuremath{\langle {\rm d}N_{\rm ch}/{\rm d}\eta \rangle_{|\eta|<1}} \\
    \hline  \xrowht[()]{10pt}
    \phantom{0}0--1\%  & 33.29$^{+0.57}_{-0.51}$  \\ \xrowht[()]{10pt}
    \phantom{0}1--5\%  & 23.44$^{+0.37}_{-0.33}$  \\  \xrowht[()]{10pt}
    \phantom{0}5--10\% & 18.25$^{+0.28}_{-0.25}$  \\  \xrowht[()]{10pt}
    10--15\% & 15.13$^{+0.22}_{-0.20}$  \\  \xrowht[()]{10pt}
    15--20\% & 12.88$^{+0.19}_{-0.17}$ \\  \xrowht[()]{10pt}
    20--30\% & 10.50$^{+0.16}_{-0.14}$ \\  \xrowht[()]{10pt}
    30--40\% &  \phantom{0}8.18$^{+0.12}_{-0.11}$  \\  \xrowht[()]{10pt}
    40--50\% &  \phantom{0}6.30$^{+0.10}_{-0.09}$  \\  \xrowht[()]{10pt}
    50--70\% &  \phantom{0}4.13 $\pm$ 0.06 \\ \xrowht[()]{10pt}
    70--100\% &  \phantom{0}1.87 $\pm$ 0.04 \\
    \hline \hline
    \xrowht[()]{6pt}
    MB values &  \phantom{0}7.47 $\pm$ 0.11 \\
    \hline
    \end{tabular}
    \end{minipage}
 \begin{minipage}[t]{0.48\linewidth}\centering
     \begin{tabular}{|c|c|}
    \hline
    Multiplicity  & p--Pb 8.16 TeV \\
    \xrowht[()]{6pt}
    class & \ensuremath{\langle {\rm d}N_{\rm ch}/{\rm d}\eta \rangle_{-1.465<\eta<0.535}}  \\
    \hline  \xrowht[()]{10pt}
    \phantom{0}0--1\%  & 73.40 $\pm$ 1.95 \\ \xrowht[()]{10pt}
    \phantom{0}1--5\%  & 56.00 $\pm$ 1.46 \\  \xrowht[()]{10pt}
    \phantom{0}5--10\% & 45.99 $\pm$ 1.20 \\  \xrowht[()]{10pt}
    10--15\% & 39.93 $\pm$ 1.03 \\  \xrowht[()]{10pt}
    15--20\% & 35.50 $\pm$ 0.93 \\  \xrowht[()]{10pt}
    20--30\% & 30.35 $\pm$ 0.79 \\  \xrowht[()]{10pt}
    30--40\% & 24.84 $\pm$ 0.65 \\  \xrowht[()]{10pt}
    40--50\% & 20.13 $\pm$ 0.53 \\  \xrowht[()]{10pt}
    50--60\% & 15.89 $\pm$ 0.31 \\ \xrowht[()]{10pt}
    60--70\% & 11.81 $\pm$ 0.41 \\  \xrowht[()]{10pt}
    70--100\% & \phantom{0}4.82 $\pm$ 0.15 \\
    \hline \hline
    \xrowht[()]{6pt}
    MB values &  20.79 $\pm$ 0.61 \\
    \hline
    \end{tabular}
    \end{minipage}

    \label{tab:1}
\end{table}
In addition, for p--Pb collisions a centrality selection based on ZN energy, that is the most unbiased selection available~\cite{ALICEcentrpA}, was used to compare data at the two available centre-of-mass energies, $\sqrt{s_{\rm{NN}}}=$~5.02 and 8.16 TeV.

Data from pp collisions are compared with MC simulations using PYTHIA~6 (Perugia 2011 tune)~\cite{PYTHIA6}, PYTHIA~8 (Monash 2013 tune)~\cite{PYTHIA8} and EPOS~LHC~\cite{EPOS} as event generators. The GEANT~3~\cite{GEANT3} particle transport code was used to track particles through the ALICE experimental set-up.
PYTHIA~6.4 Perugia 2011 has a particular tune of the initial, final state showers and UE modelling. In particular, the tuning of the beam remnant is done using MB observables such as charged-particle multiplicity and transverse momentum spectra measured by CDF. In this tune, some early LHC data were used for the tuning of beam remnants.
PYTHIA~8 Monash is the default tune of PYTHIA~8.1 which uses UE data from the LHC, SPS and Tevatron to tune the parameters relevant for multi-partonic interactions (MPI). No particular tune of the beam remnant is implemented.
EPOS~LHC uses early LHC data for tuning.
The particle production has two main components: the strings composed of pomerons at midrapidity, while the remnants carry the remaining energy covering mostly the fragmentation region.

\section{Forward--backward energy asymmetry in pp collisions}\label{sec.4}

\begin{figure}[ht]
    \begin{center}
    \includegraphics[width = 0.48\textwidth]{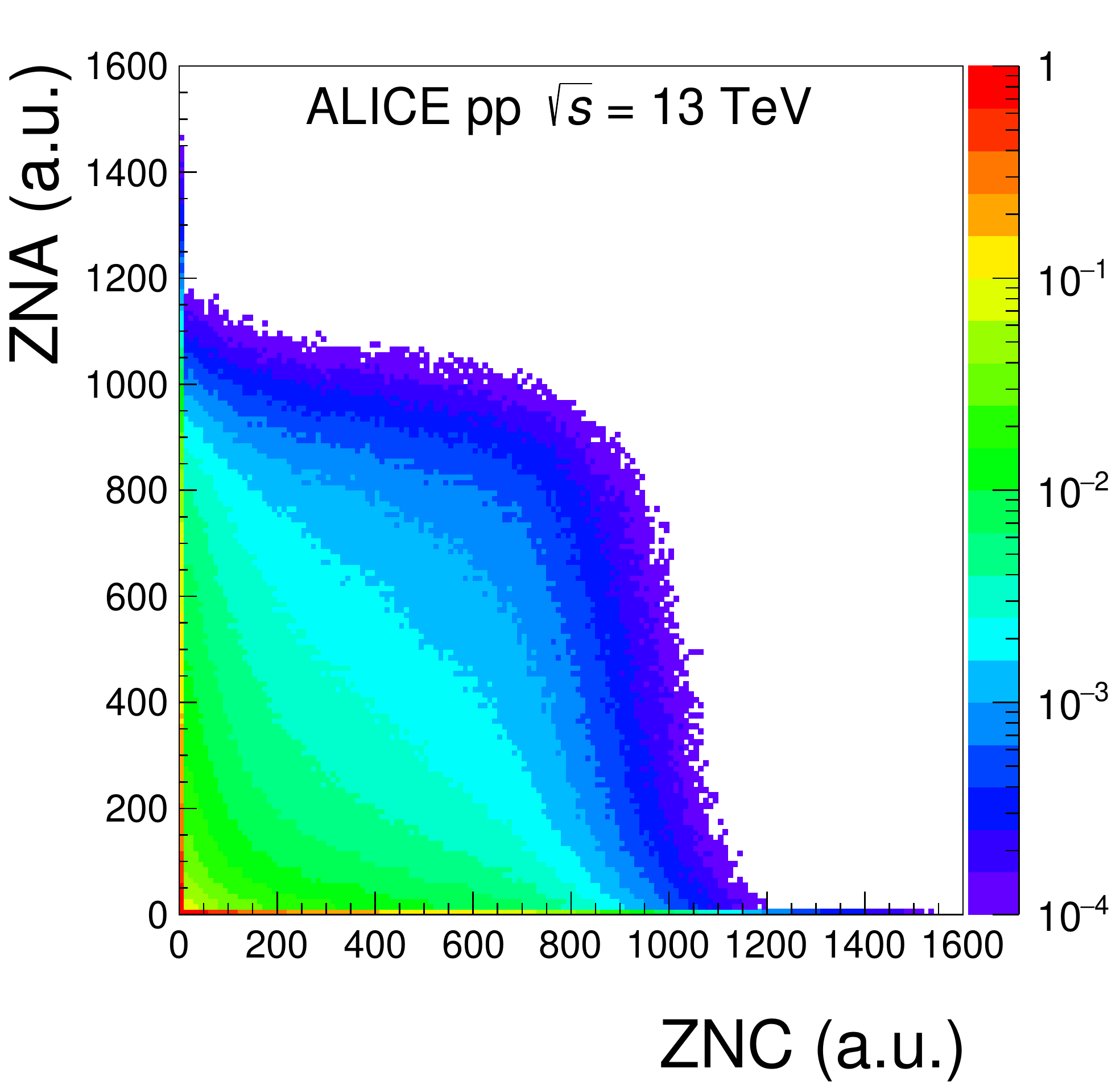}
    \includegraphics[width = 0.48\textwidth]{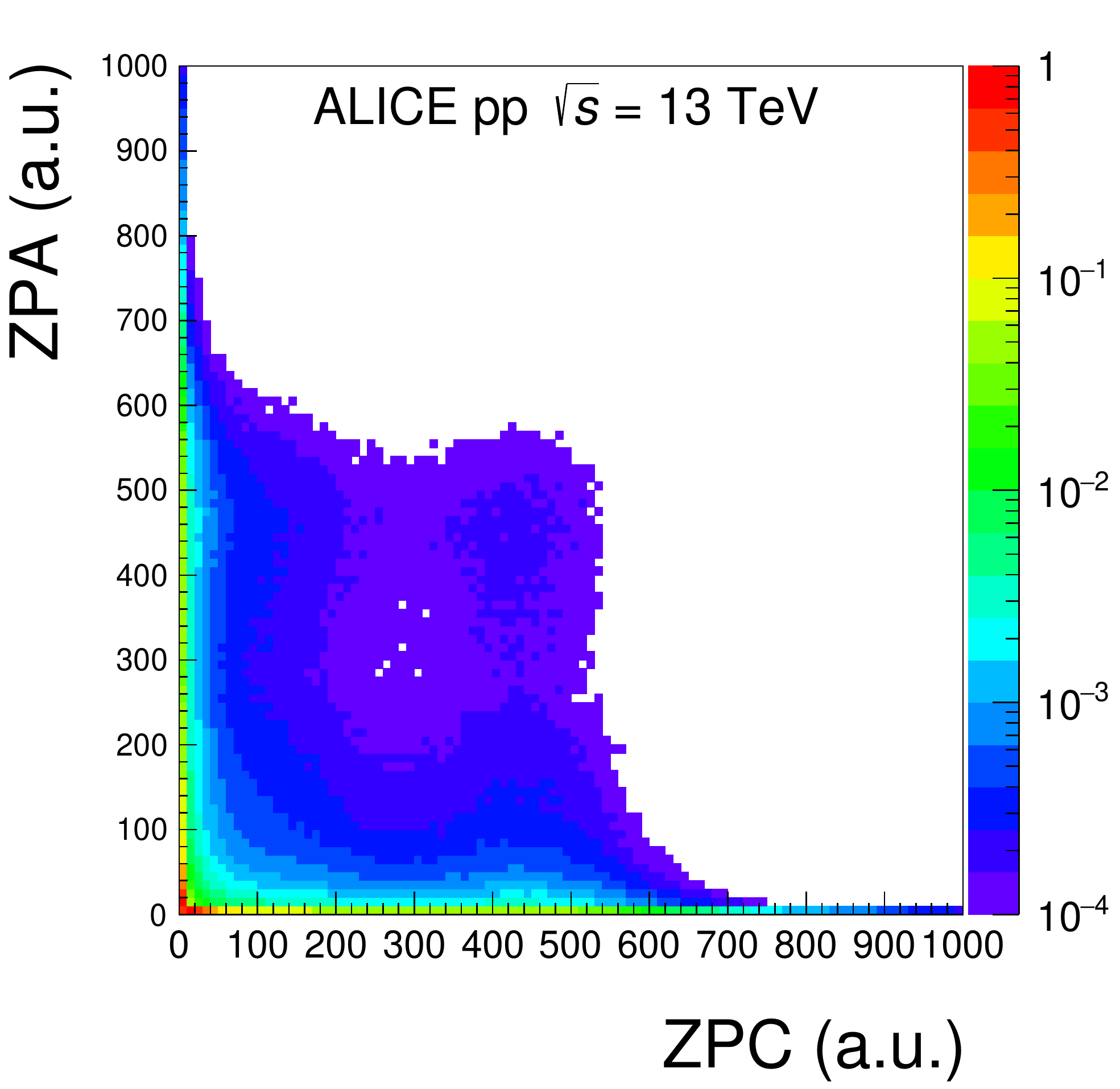}
    \end{center}
    \caption{A- vs. C-side signal in ZN (left) and ZP (right). The maximum at (0,0) is normalised to unity, while the minimum is limited to $10^{-4}$.}
    \label{fig:1}
\end{figure}

The emission of very forward energy can be investigated over a pseudorapidity gap of more than 18 units using both ZDC systems, placed on the two sides relative to the IP.
The correlation between the energy emitted at forward and backward rapidities in MB collisions is shown in  Fig.~\ref{fig:1}, separately for ZN and ZP. For both detectors the signals increase linearly with the detected energy, without showing saturation or non-linear effects for large deposited energies.
In both cases, the detected energies show asymmetric features, namely a high  energy deposit on one side corresponds to a very low energy deposition on the opposite side. In addition, a subset of the event sample is correlated, in particular for neutron emission.
The origin of the observed asymmetry in energy at forward and backward rapidities was further investigated in MC generators, without considering any tracking through the experimental set-up, but selecting solely on particle kinematics. It was found that the energy emitted at large rapidities becomes asymmetric once the phase space is restricted to the pseudorapidity ranges covered by ZN and ZP detectors. In conclusion, the observed asymmetry is an effect due to the limited and not overlapping pseudorapidity ranges covered by the neutron and the proton detectors.

\begin{figure}[ht]
    \begin{center}
    \includegraphics[width = 0.9\textwidth]{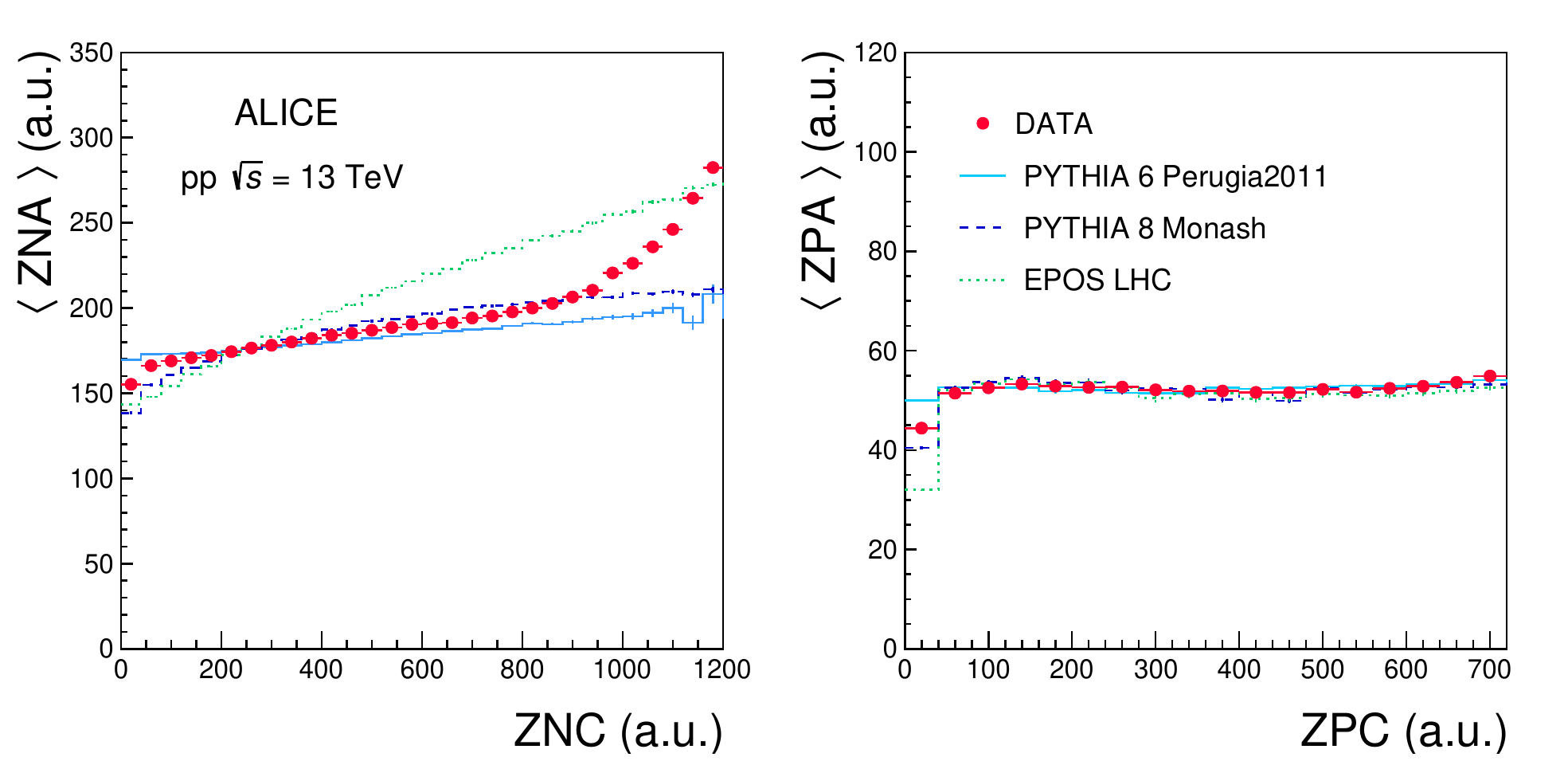}
    \end{center}
    \caption{Average A-side ZN (left) and ZP (right) signals as a function of the C-side signals in pp collisions at $\sqrt{s} = $~13~TeV. Data (red full circles) are compared with model predictions from PYTHIA~6 (azure line), PYTHIA~8 (dashed blue line) and EPOS (dotted green line).}
    \label{fig:3}
\end{figure}

Another way to gain insight into the correlation between energy emitted at backward and forward rapidities is to study the average signal on one side as a function of the signal on the other side. Results are shown in Fig.~\ref{fig:3}, together with comparisons to model calculations. The energies carried by leading protons in the two sides are not correlated, as was already observed at lower energies at the ISR~\cite{Basile}.
In contrast to this, the data indicate the presence of a correlation between the energies carried by leading neutrons on opposite sides, in particular at high energies. According to all the models used for comparison, the background contribution from photons is relevant only at very low energies (up to ~1 TeV), while for higher energies the signal is only due to leading neutrons.
Models predict a flat behaviour for leading protons in good agreement with data, while for neutrons, even though they show some degree of correlation between the neutron energies, none of them is able to reproduce quantitatively the measured dependence over the whole range.

\section{Forward energy as a function of charged-particle production at midrapidity}\label{sec.5}

From phenomenological models, one expects that selecting events characterised by a larger than average multiplicity, or by the emission of a large transverse momentum particle, corresponds to selecting collisions with a smaller than average impact parameter~\cite{ms:1, ms:3}, and a larger than average number of MPI~\cite{skands}.
The forward energy detected by the ZDCs was studied as a function of the charged-particle multiplicity produced at midrapidity (in $|\eta|<1$ for pp collisions at $\sqrt{s}=13$~TeV and $-1.465<\eta<0.535$ for p--Pb collisions at $\sqrt{s_{\rm{NN}}}$~=~8.16~TeV), and as a function of the leading particle transverse momentum, $\pt^{\rm leading}$,  in $|\eta|<0.8$ in pp collisions.
To study the interaction of a proton either with another proton or with a Pb nucleus, the dependence of the very forward energy on charged particle production at midrapidity was studied in both pp and p--Pb collisions.

\subsection{Very forward energy in p--Pb collisions}\label{sec5.0}

\begin{figure}[ht]
    \begin{center}
    \includegraphics[width = 0.48\textwidth]{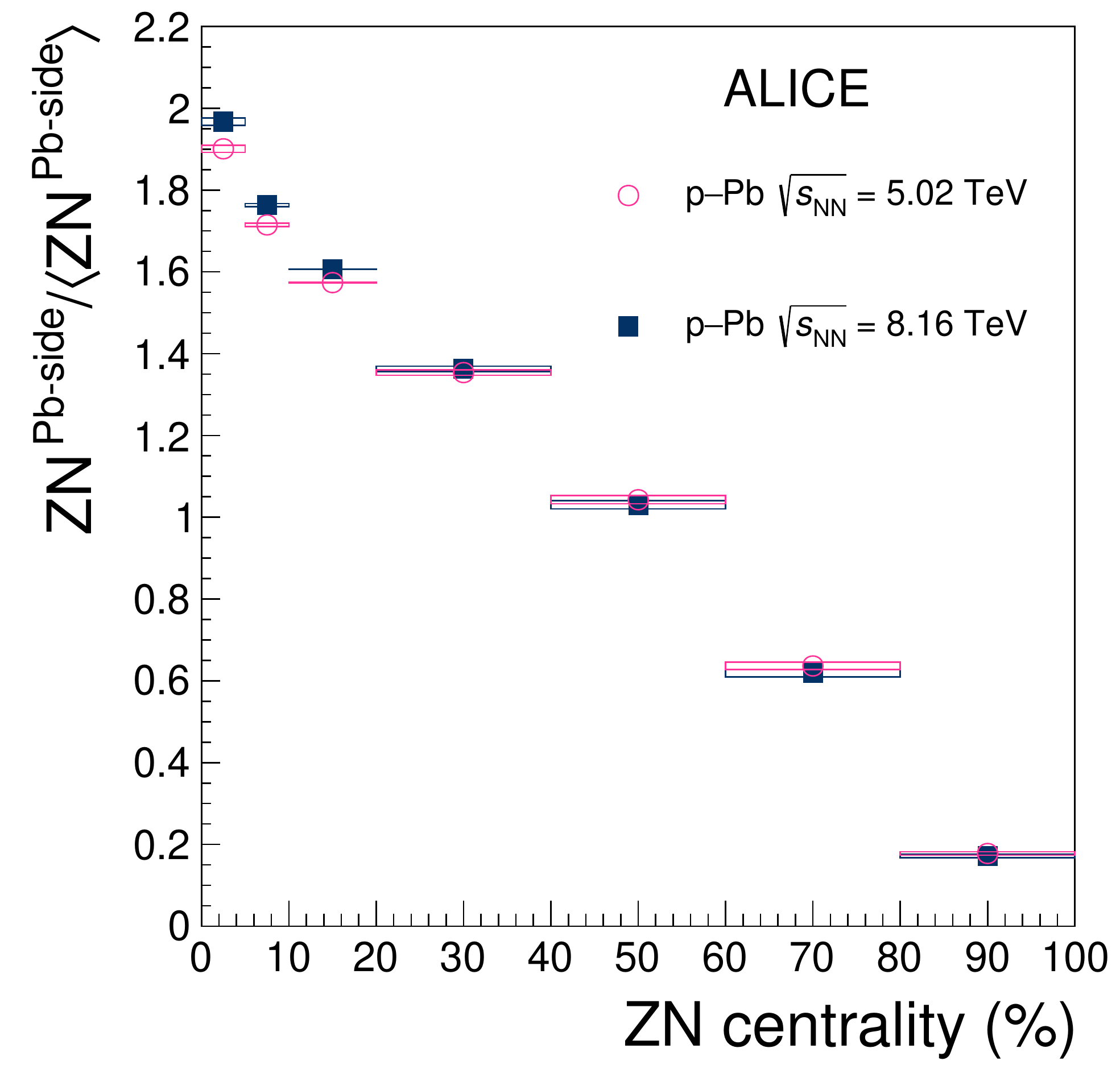}
    \includegraphics[width = 0.48\textwidth]{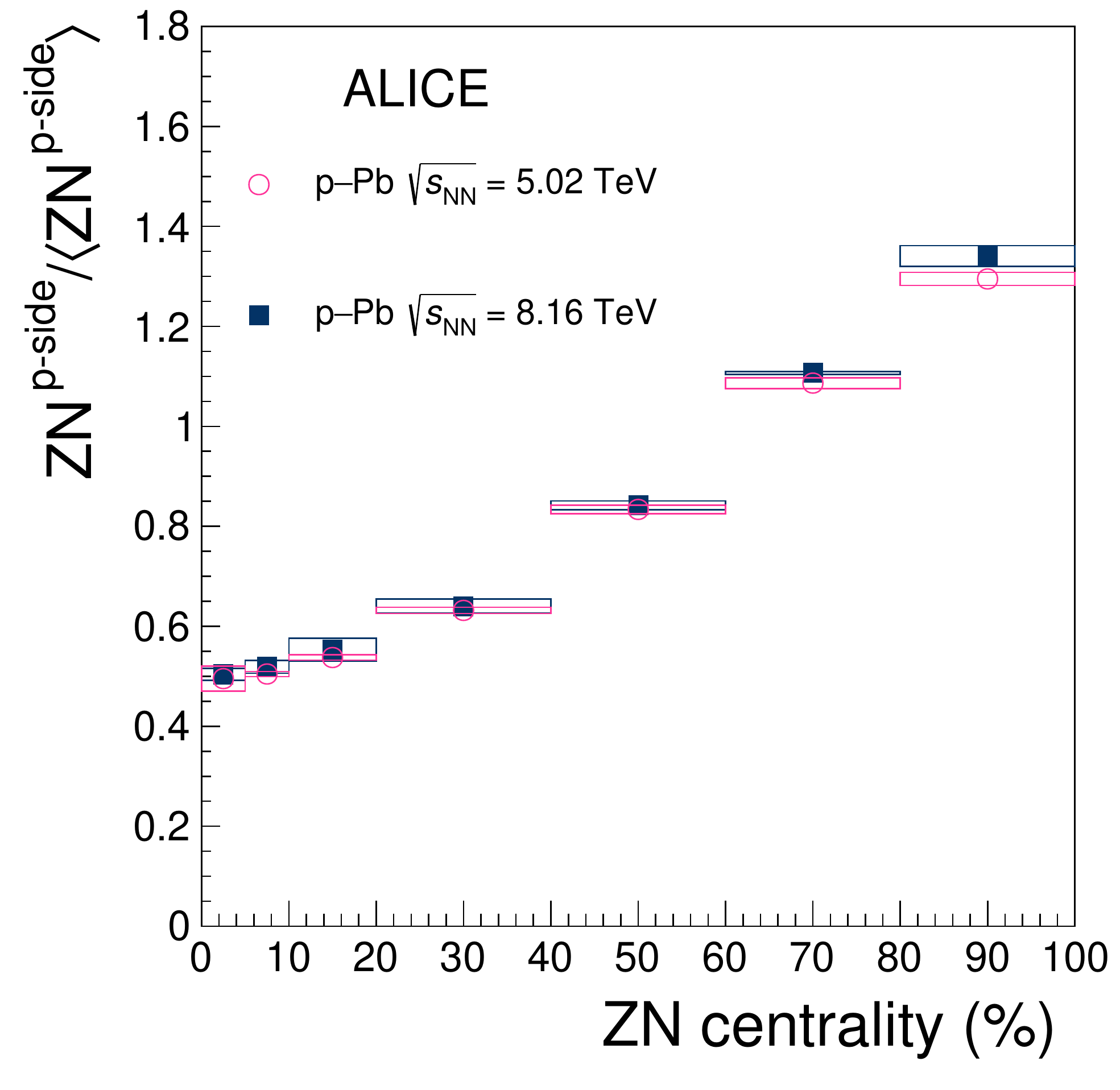}
    \end{center}
    \caption{ZN energy normalised to the average MB value in the Pb-fragmentation (left) and in the p-fragmentation (right) regions as a function of centrality estimated from ZN~\cite{ALICEcentrpA} in p--Pb collisions at $\sqrt{s_{\rm{NN}}}=5.02$~TeV (pink circles) and $8.16$~TeV (blue squares). The boxes represent the systematic uncertainty.}
    \label{fig:5}
\end{figure}

In p--Pb collisions, the p-fragmentation and the Pb-fragmentation sides show a complementary behaviour as a function of centrality, as already described for $\sqrt{s_{\rm{NN}}}=$~5.02~TeV p--Pb collisions~\cite{ALICEcentrpA}.
In Fig.~\ref{fig:5}, the self-normalised ZN signals as a function of centrality, estimated through the energy measured by the neutron calorimeter in the Pb-fragmentation region as described in Ref.~\cite{ALICEcentrpA}, are shown for the Pb-fragmentation and for the p-fragmentation sides. Events characterised by a large multiplicity (corresponding to central events) have a large forward energy deposit in the Pb-fragmentation side and a small energy deposit in the p-fragmentation side. This behaviour does not show a strong dependence on the collision energy.
\begin{figure}[ht]
    \begin{center}
    \includegraphics[width = 0.48\textwidth]{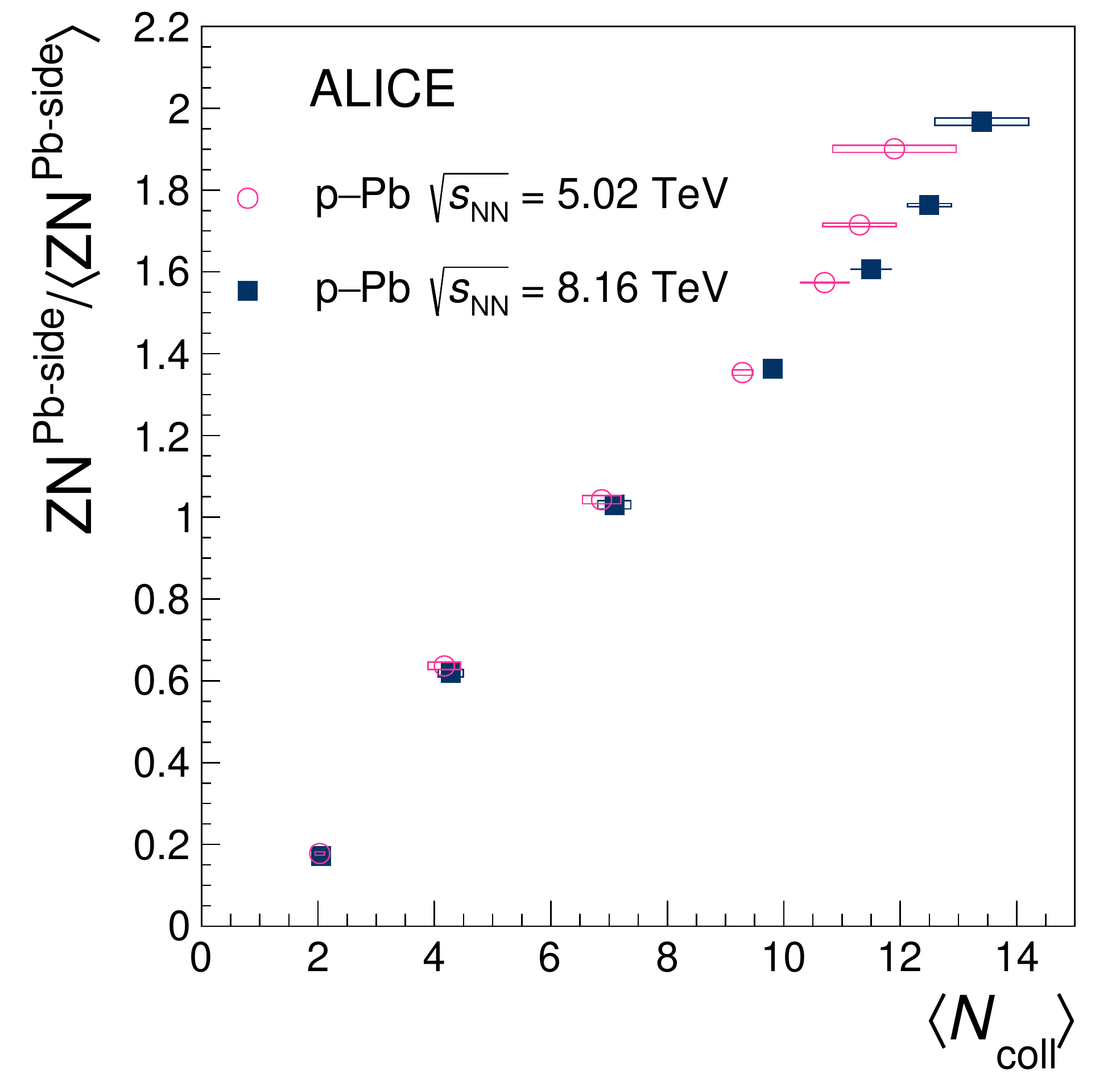}
    \includegraphics[width = 0.48\textwidth]{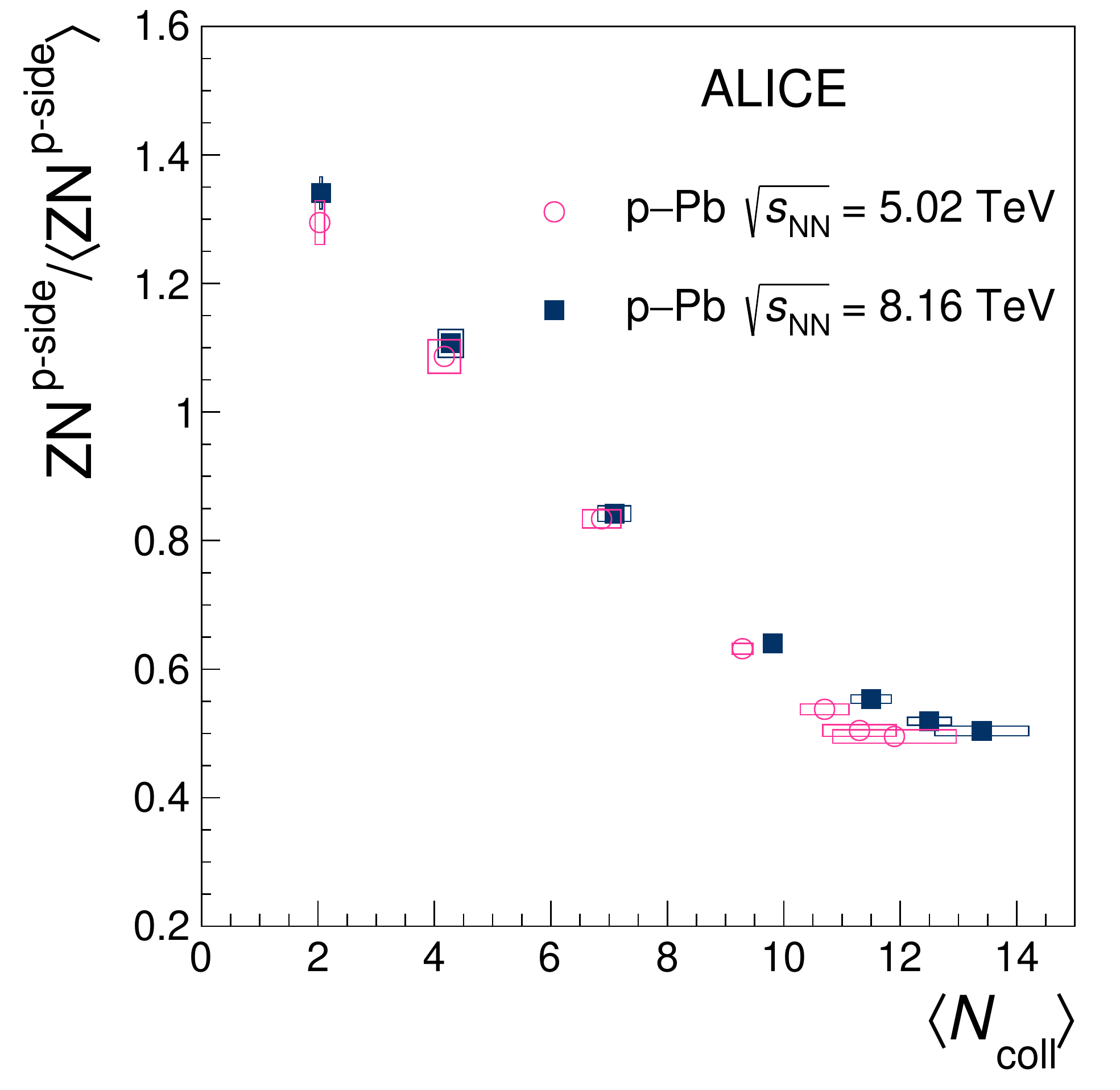}
    \end{center}
    \caption{ZN energy normalised to the average MB value in the Pb-fragmentation (left) and in the p-fragmentation (right) regions as a function of the average $N_{\rm coll}$ in p--Pb collisions at $\sqrt{s_{\rm{NN}}}=5.02$~TeV (pink circles) and $8.16$~TeV (blue squares). The boxes represent the systematic uncertainty.}
    \label{fig:5.5}
\end{figure}

In Fig.~\ref{fig:5.5}, the normalised energies in the two fragmentation regions are shown as a function of the number of binary nucleon--nucleon collisions, $N_{\rm coll}$, calculated as described in Ref.~\cite{ALICEcentrpA}. The $N_{\rm coll}$ values are included in Ref.~\cite{ANcentralita} for both p--Pb colliding energies. It is interesting to notice how the very forward energy in the p-fragmentation region is, not only inversely dependent on centrality, but also decreases linearly with the number of binary collisions over a wide range of centralities.

\subsection{Forward energy dependence on charged-particle multiplicity at midrapidity}\label{sec5.1}

\begin{figure}[ht]
    \begin{center}
    \includegraphics[width = 0.96\textwidth]{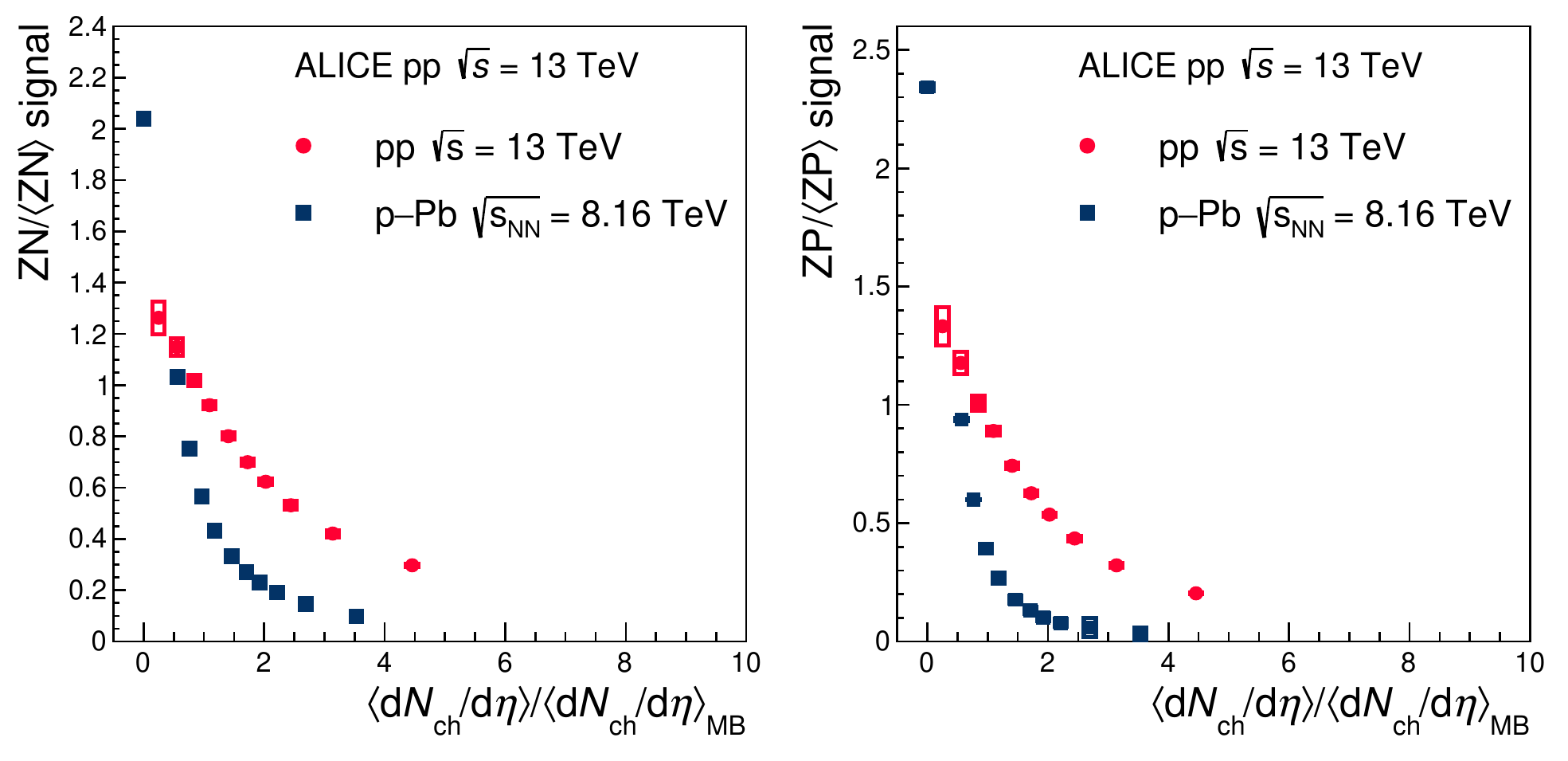}
    \end{center}
    \caption{ZN (left) and ZP (right) self-normalised signals as a function of the normalised multiplicity measured in 2 units of $\eta$ around the centre-of-mass midrapidity in pp (red circles) collisions and in the p-fragmentation region in p--Pb (blue squares) collisions. The boxes represent systematic uncertainties.}
    \label{fig:4}
\end{figure}

The normalised ZN and ZP signals  are compared, as a function of the self-normalised charged-particle multiplicity measured at midrapidity in pp and in the p-frag\-men\-ta\-tion region in p--Pb collisions, in Fig.~\ref{fig:4}.
In pp collisions the ZDC self-normalised values are averaged between the two p-fragmentation sides, while for p--Pb collisions two different data taking periods with inverted beam directions were averaged.

For pp collisions, two sources of systematic uncertainty were considered: the trigger selection and the difference between the measurements performed on both sides.
The first contribution was estimated using a different trigger selection based on the AD detector, which removed some residual contribution from single-diffractive events (estimated to be below 3\textperthousand \  at 8~TeV\cite{ALICEppmult-SD}). This uncertainty ranges from $2\%$ to $5\%$ for ZN and from $2\%$ to $6\%$ for ZP.
The uncertainty coming from considering the energy measured in two sides ranges from $0.3\%$ to $1\%$ for ZN and from $0.1\%$ to $1\%$ for ZP.
The total uncertainty is calculated as the sum in quadrature of the two contributions, and ranges from 2\% to 5\% for ZN and from 2\% to 6\% for ZP.
In p--Pb collisions, the difference between the two beam configurations  was considered as a source of systematic uncertainty, and it ranges from $0.4\%$ to $4\%$ for ZN and $0.1\%$ to $28\%$ for ZP. The higher uncertainty values correspond to higher multiplicity bins, where the ZP signal is small, and this leads to a small absolute uncertainty value.
The dependence of the detected forward energy on midrapidity multiplicity shows similar features in pp and in the p-fragmentation region in p--Pb collisions: the higher is the activity measured at midrapidity, the smaller is the forward energy.

\begin{figure}[ht]
    \begin{center}
    \includegraphics[width = 0.48\textwidth]{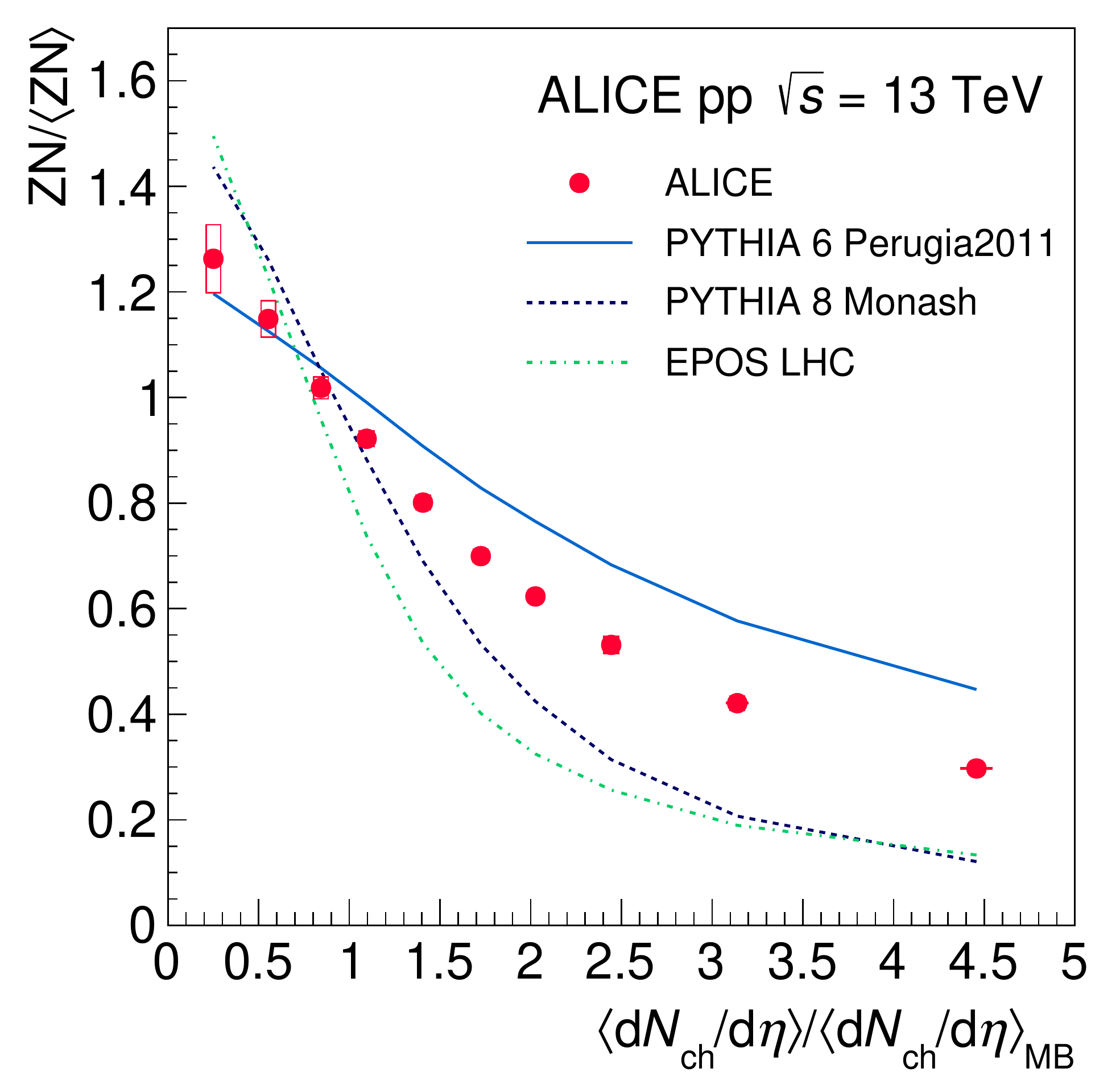}
    \includegraphics[width = 0.48\textwidth]{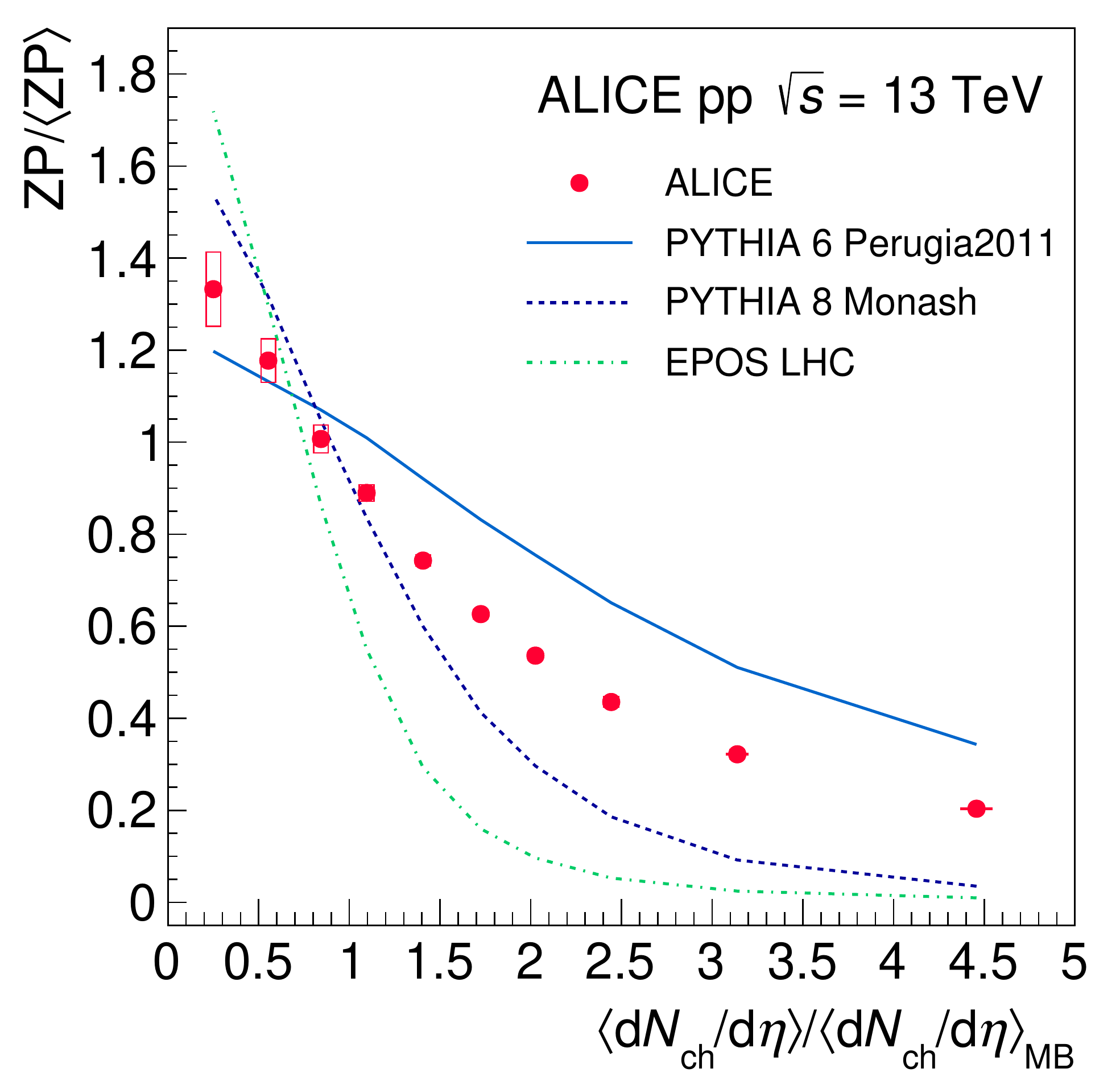}
    \end{center}
    \caption{Self-normalised ZN (left) and ZP (right) signals as a function of the normalised charged-particle multiplicity produced in $|\eta|<1$ in pp collisions. Data (red markers) are compared with PYTHIA~6 (blue solid line), PYTHIA~8 (blue dashed line) and EPOS~LHC (green dotted line). }
    \label{fig:6}
\end{figure}

The self-normalised forward energy as a function of the average multiplicity in a certain interval, normalised to the MB average, $\langle \rm{d}N/\rm{d}\eta \rangle/\langle \rm{d}N/\rm{d}\eta \rangle_{\rm{MB}}$, was compared with MC simulations for pp collisions. All models are able to describe the overall decreasing trend, and PYTHIA 6 (Perugia 2011) is the one showing the best degree of agreement, as shown in Fig.~\ref{fig:6}. However, none of the models is able to reproduce the experimental results quantitatively. Moreover, they are not able to satisfactorily describe the measured forward energy spectra in multiplicity bins.

\begin{figure}[tb]
    \begin{center}
    \includegraphics[width = 0.48\textwidth]{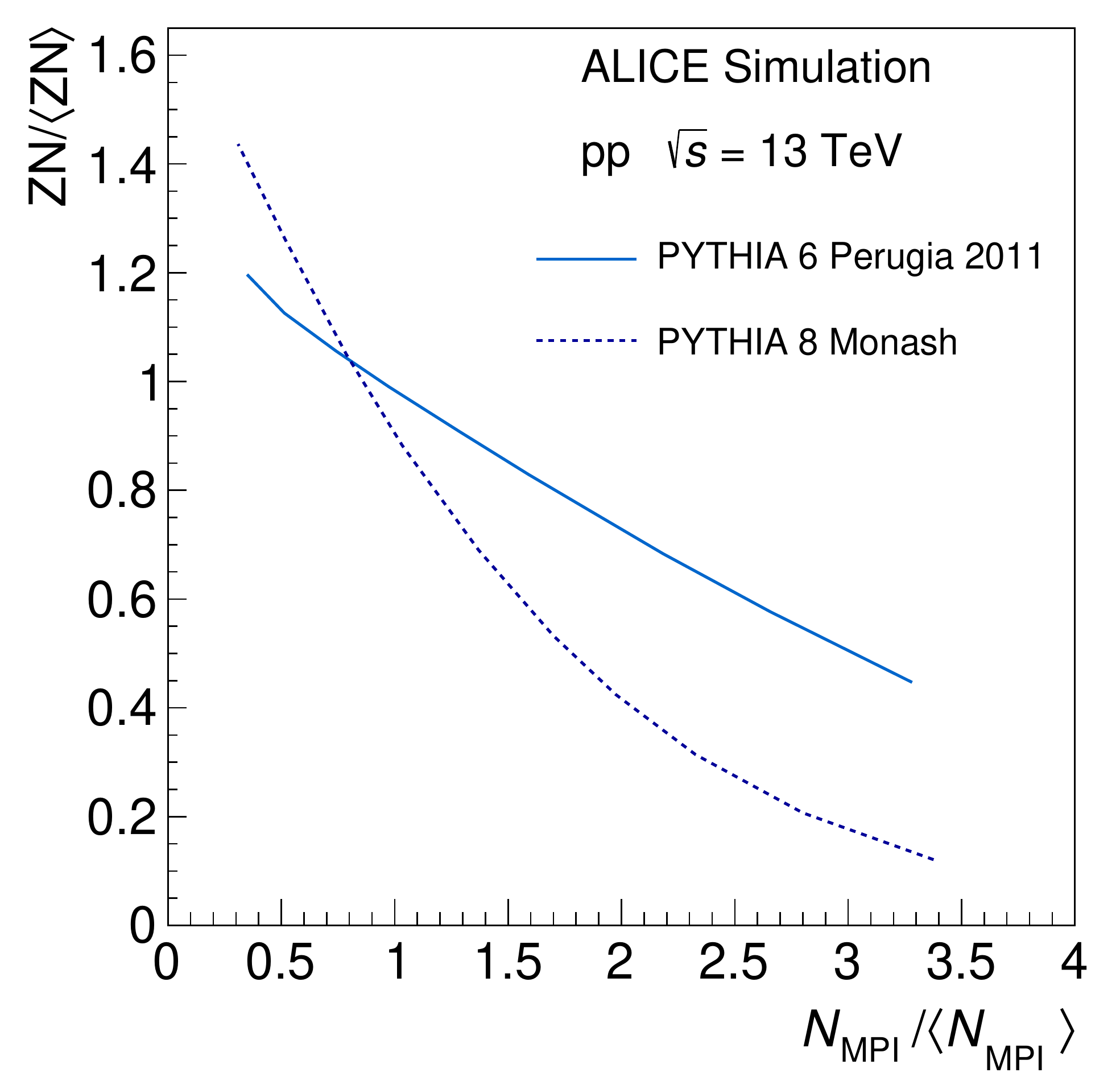}
    \includegraphics[width = 0.48\textwidth]{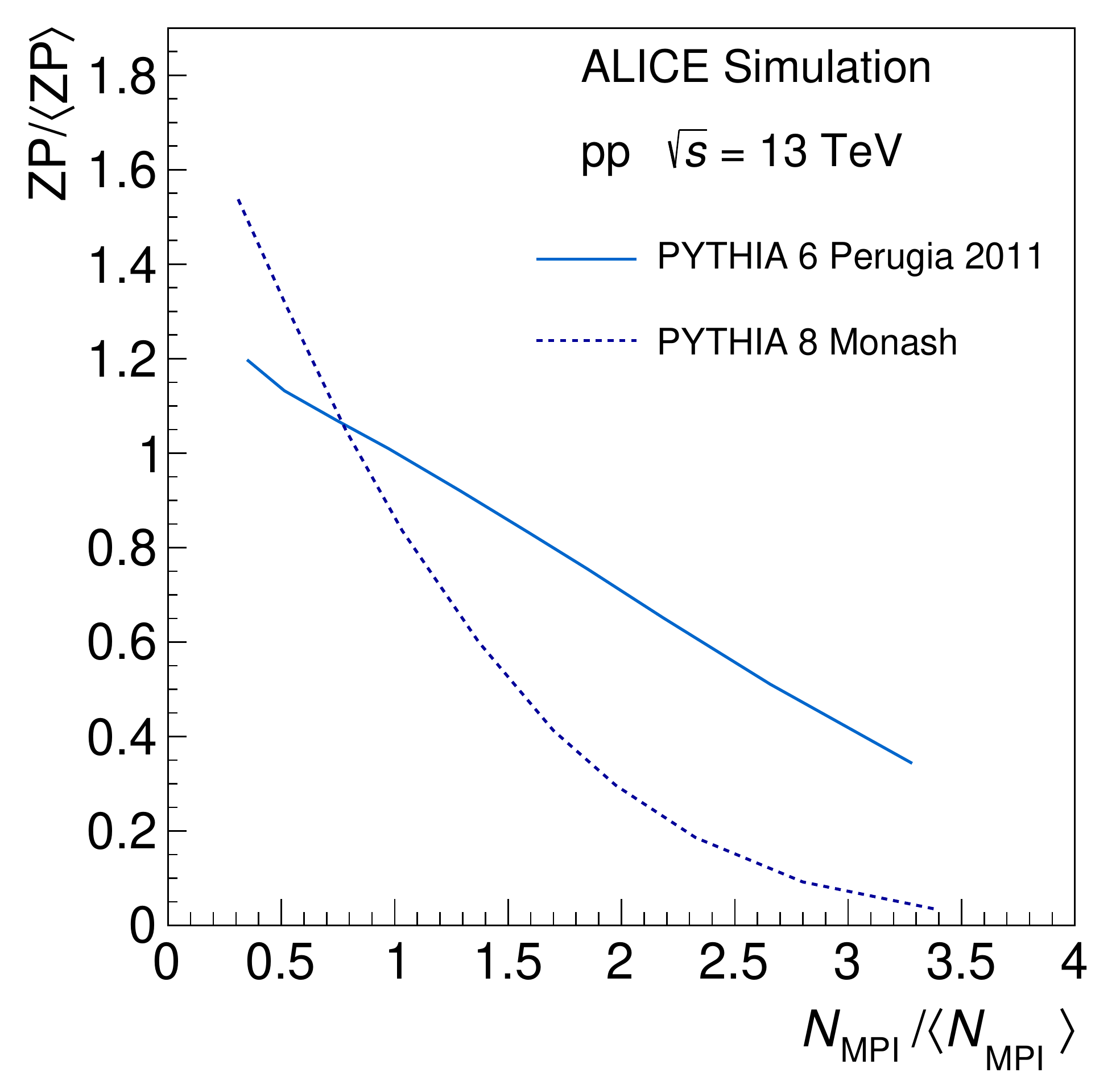}
    \end{center}
    \caption{Self-normalised ZN (left) and ZP (right) signal as a function of the number of self-normalised MPI extracted from PYTHIA~6 Perugia 2011 (solid line) and PYTHIA~8 Monash (dashed line) tunes.}
    \label{fig:11}
\end{figure}

The PYTHIA event generator includes MPI modelling, which is needed to reproduce the energy flow measurements at LHC energies at smaller rapidities~\cite{CMS, CMSEflow}. The correlation between the very forward energy and the number of MPIs was studied using PYTHIA.
Figure~\ref{fig:11} shows the simulated ZN and ZP self-normalised responses as a function of the number of MPIs, normalised to their MB average value ($\langle N_{\rm MPI} \rangle = $~4.0 for PYTHIA~6 Perugia 2011 and $\langle N_{\rm MPI} \rangle = $~5.0 for PYTHIA~8 Monash). Both models predict a clear relationship between the very forward energy and $N_{\rm MPI}$, showing a decrease in the average very forward energy for an increasing number of MPIs, albeit with different slopes due to the different treatment of MPIs in the two generators. This pattern resembles the observed dependence on charged-particle multiplicity (see Fig.~\ref{fig:6}), as can be expected in an impact-parameter dependent MPI picture~\cite{mpiSjo}.

\begin{figure}[ht]
    \begin{center}
    \includegraphics[width = 0.48\textwidth]{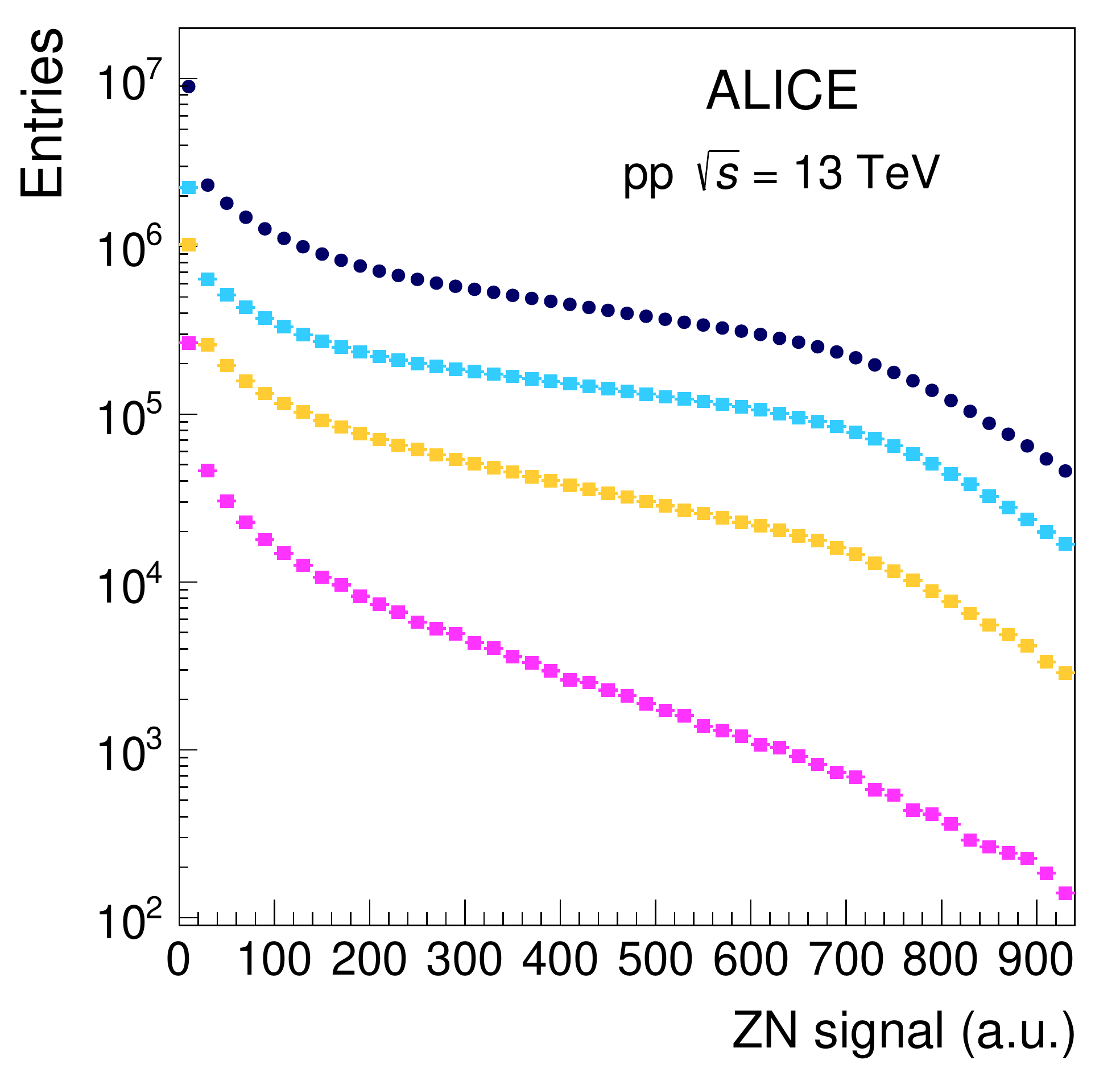}
    \includegraphics[width = 0.48\textwidth]{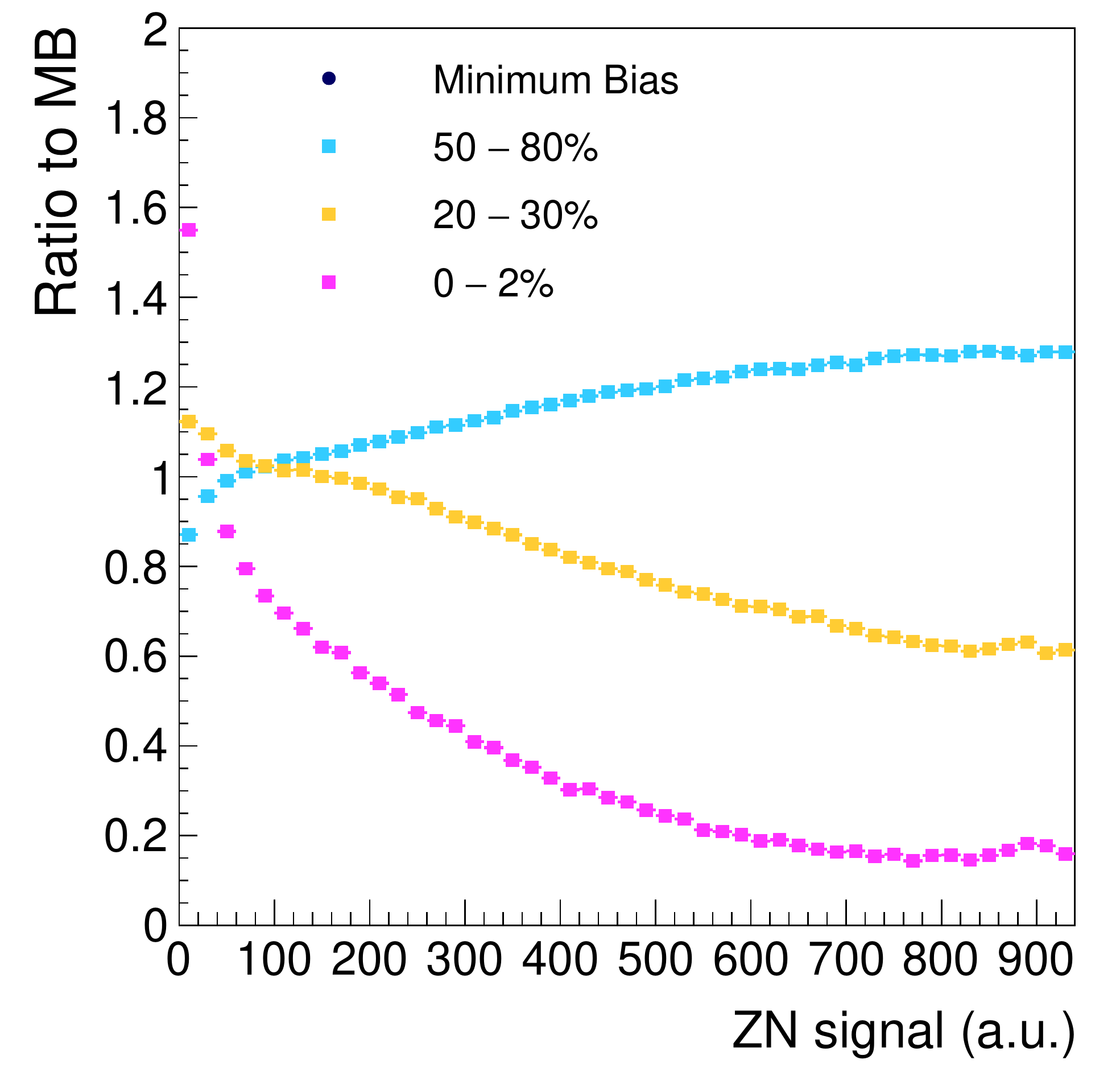}
    \end{center}
    \caption{Left: ZN spectrum in pp collisions at $\sqrt{s}=$13~TeV for the MB sample (blue circles) and in three multiplicity intervals: high (magenta squares), intermediate (orange squares) and low (azure squares) multiplicity. Right: ratio of the spectra, normalised to the number of events in each bin, in the three multiplicity intervals to the MB spectrum.}
    \label{fig:10}
\end{figure}

The shape of the ZN spectrum also showed a dependence on the charged-particle multiplicity at midrapidity. To characterise these modifications, three narrow multiplicity intervals were selected, corresponding to high (0--2\%), intermediate (20--30\%) and low (50--80\%) multiplicities, and the ZN spectrum was compared to the MB distribution in these intervals.
Figure~\ref{fig:10} shows the spectrum modifications in the considered multiplicity intervals. In particular, as predicted in Refs.~\cite{ms:3, ms:4}, the forward leading neutron energy is suppressed when a higher activity is measured at midrapidity.

\subsection{Correlation between forward energy and leading particle transverse momentum}\label{5.2}

\begin{figure}[ht]
    \begin{center}
    \includegraphics[width = 0.48\textwidth]{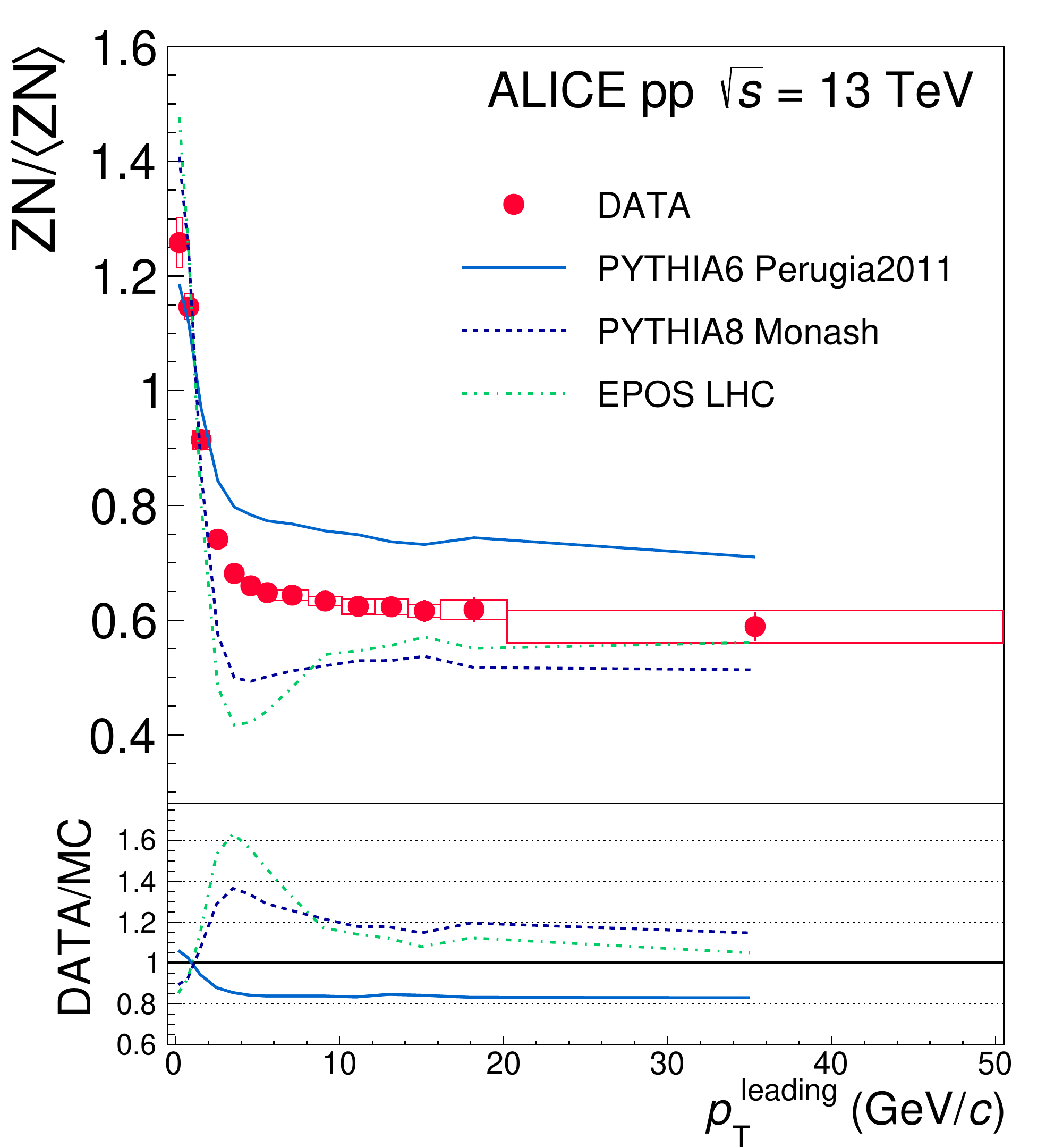}
    \includegraphics[width = 0.48\textwidth]{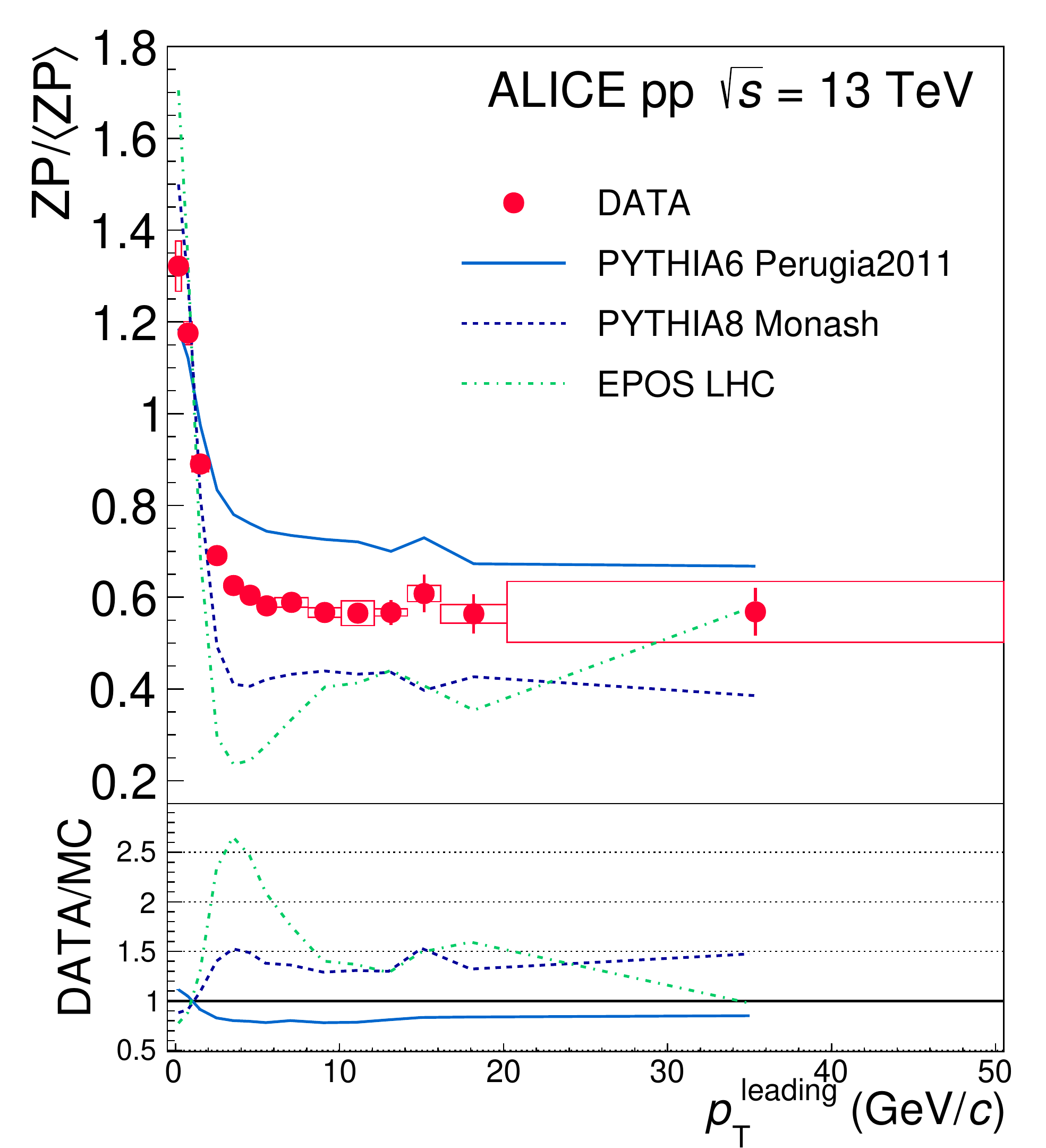}
    \end{center}
    \caption{Self-normalised ZN (left) and ZP (right) signals as a function of the leading particle \pt measured in $|\eta|<0.8$ in pp collisions at $\sqrt{s}=$13~TeV. Data (red markers) are compared to PYTHIA~6 (blue solid line), PYTHIA~8 (blue dashed line) and EPOS~LHC (green dotted line).}
    \label{fig:7}
\end{figure}

The forward energy was studied as a function of the leading particle transverse momentum, $\pt^{{\rm leading}}$, defined event by event as the track with the largest transverse momentum in $|\eta|<0.8$. Events with large forward energies are characterised by smaller values of the leading particle \pt, as already observed in measurements at smaller pseudorapidities at the LHC~\cite{ATLAS}.
The self-normalised ZN and ZP signals as a function of the leading particle \pt are shown in Fig.~\ref{fig:7}.
The total systematic uncertainty was estimated as the sum in quadrature of three contributions: the first one comes from the trigger selection, the second is due to the differences between the measurements performed on the two sides, and the third contribution takes into account the misidentification of the leading particle, which is corrected using a data-driven procedure, as detailed in Ref.~\cite{ALICEue0}. The total uncertainty ranges from 0.8\% to 4.4\% for ZN and from 0.8\% to 5.5\% for ZP, the dominant contribution coming from the difference between the two sides.
A large forward energy is measured for very low values of the leading particle \pt ($\pt^{{\rm leading}}<1$~GeV/$c$). For larger leading particle \pt values,  ZN and ZP energies rapidly decrease and saturate for $\pt^{{\rm leading}} \gtrsim 5$~GeV/$c$.
The same trend was reported by the CMS collaboration when measuring the energy  at smaller rapidities ($-6.6< \eta <-5.2$)~\cite{CMS}.

The self-normalised ZN and ZP signals as a function of the leading particle \pt are also compared to the three models under consideration, as shown in Fig.~\ref{fig:7}. The trend predicted by PYTHIA~6 is qualitatively in agreement with data, even if, for a leading particle \pt larger than $2$~GeV/$c$, PYTHIA~6 overestimates the forward energy by almost $20\%$. In contrast, the leading particle \pt dependence predicted by PYTHIA~8  is different for intermediate \pt values (2--8~GeV/$c$). This seems to indicate that the treatment of colour reconnections and beam remnants in the Perugia tunes of PYTHIA~6 is more realistic than the one in the default Monash tune of PYTHIA~8. The core–corona EPOS~LHC event generator also predicts a quite different pattern, in particular it shows a depletion for intermediate \pt values (2--8~GeV/$c$), where collective expansion (flow) included in the core part of the model might play a major role.

\subsection{Very forward energy and event properties in pp collisions}

\begin{figure}[ht]
    \begin{center}
    \includegraphics[width = 0.46\textwidth]{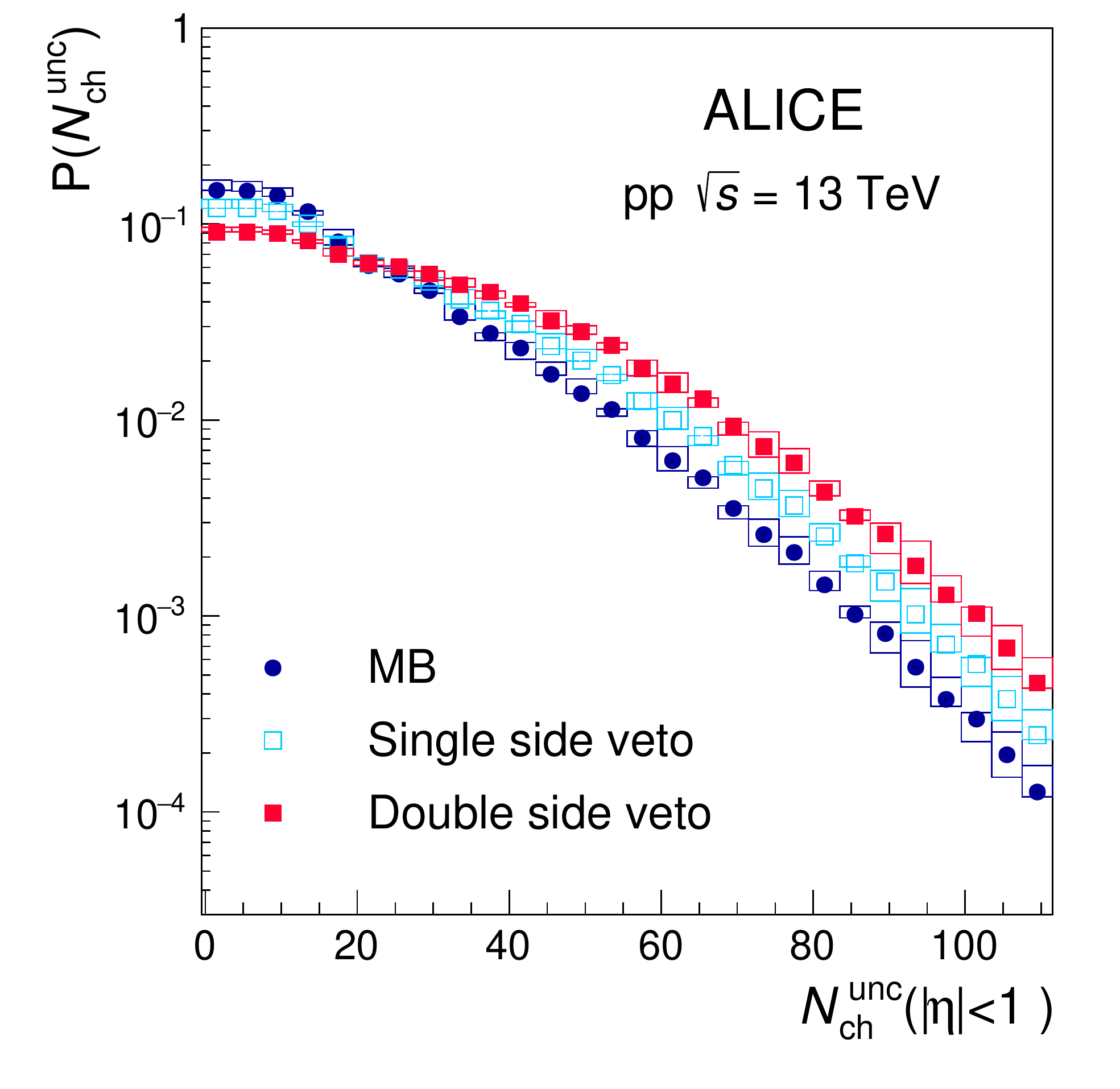}
    \includegraphics[width = 0.46\textwidth]{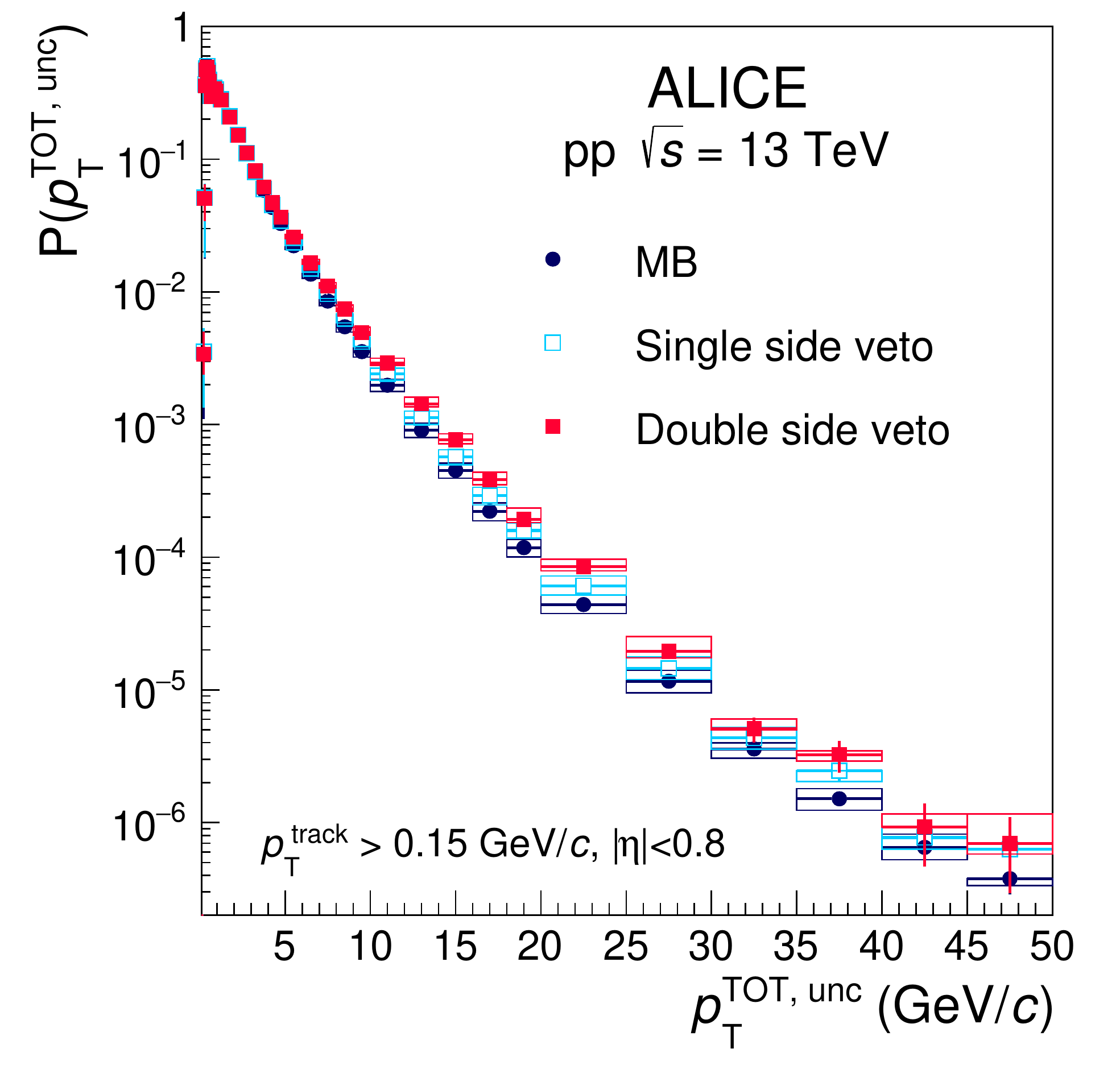}
    \end{center}
    \caption{Distributions of the (not fully corrected) number of charged particles in $|\eta|<1$ (left) and of the total transverse momentum in $|\eta|<0.8$ (right) for MB interactions (blue circles), single-side (azure empty squares) and double-side (red full squares) veto conditions on leading baryon production in pp collisions at $\sqrt{s}=$13 TeV.}
    \label{fig:8}
\end{figure}

The energy carried forward by leading baryons has been proposed as a tool to select events with smaller than average impact parameter. In Ref.~\cite{ms:2} a double veto on leading baryon production is suggested as an effective way to select events characterised by a narrower impact-parameter distribution and harder particle spectra.
To test this hypothesis, the energy detected by the ZDCs was used as a veto, requesting that neither ZN nor ZP have a signal on one or on both sides, thus defining a single-side and a double-side veto condition on leading baryon production, respectively. The charged-particle multiplicity in $|\eta|<1$ and the total transverse momentum in $|\eta|<0.8$, ${p_{\rm T}}^{\rm{TOT}}$, were compared applying these conditions.
The charged particle distributions are corrected using MC simulations with different generators, and the systematic uncertainty is estimated using EPOS LHC and PYTHIA~8. The correction factors for the total \pt distributions, to account for tracking efficiency and secondary contamination, were extracted from~\cite{ALICEue}, and the systematic uncertainty is estimated varying the \pt within the obtained boundaries.
The distributions are corrected only for inefficiencies and not for effects related to resolution. However, they provide a clear indication of the effect due to the different applied conditions.
In Fig.~\ref{fig:8}, $N_{\rm ch}$ and ${p_{\rm T}}^{\rm{TOT}}$ distributions for the MB and for the two vetoed samples are shown.
The average values of $N_{\rm ch}$ are a factor $\sim$1.2 and $\sim$1.5 higher than in the MB sample for the single and double veto selection, respectively.  The analysis of the simulated samples provided similar results, yielding to similar increases for the average values of $N_{\rm ch}$, namely a factor (1.1) 1.3 for PYTHIA~6, (1.2) 1.6 for to PYTHIA~8 and (1.2) 1.4 using EPOS~LHC, applying the (single) double veto condition.
In conclusion, a double-side veto condition selects a larger than average multiplicity and a harder \pt distribution at midrapidity. This measurement supports the prediction that a double-side veto on baryon production leads to a narrower impact-parameter distribution.

\begin{figure}[ht]
    \begin{center}
    \includegraphics[width = 0.6\textwidth]{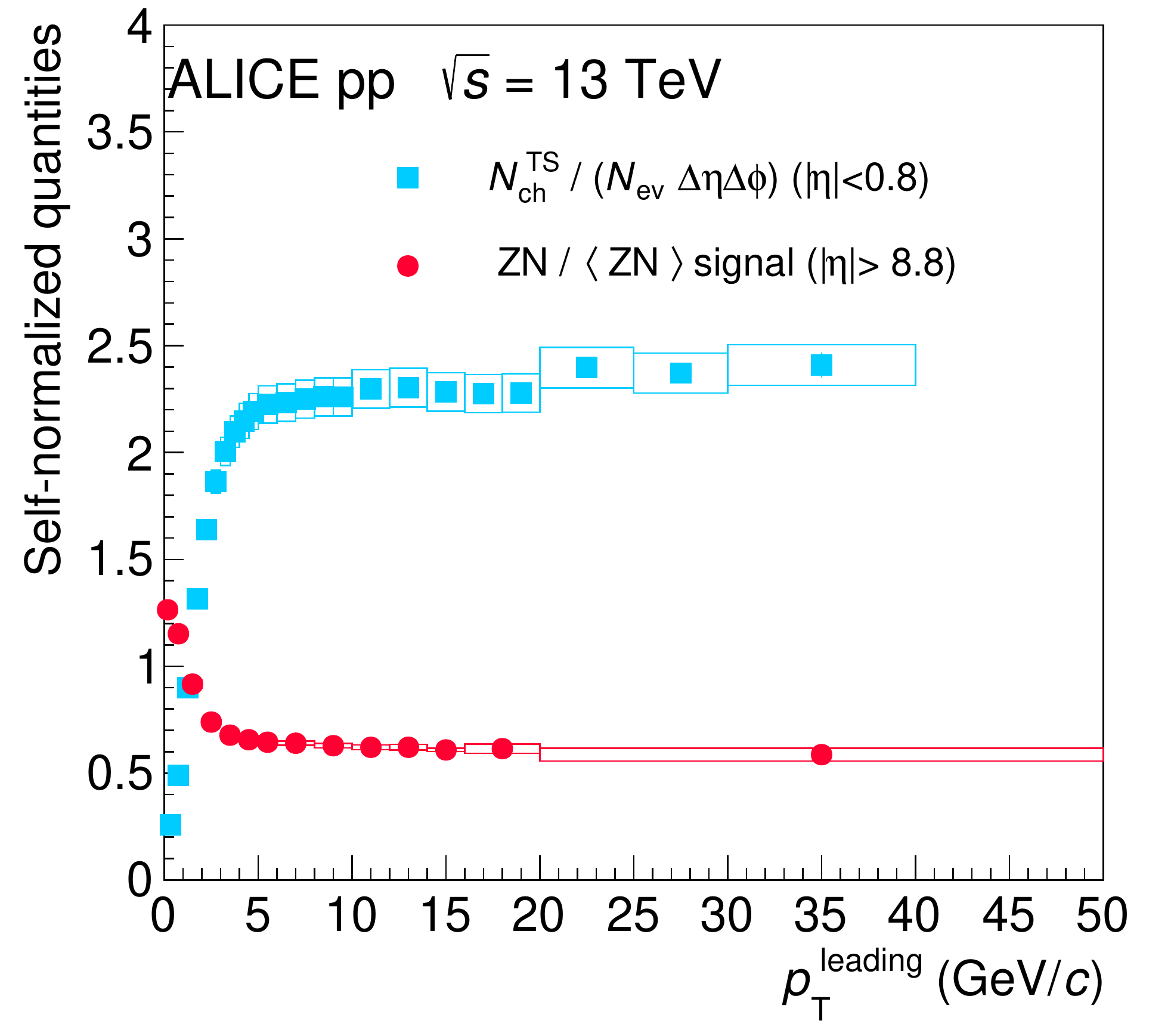}
    \end{center}
    \caption{Self-normalised ZN signal (red circles) and number density $N_{ch}$ (azure squares) distribution in the transverse region (published in Ref.~\cite{ALICEue})  as a function of $\pt^{\rm leading}$ measured in $|\eta| <$0.8. Tracks have \pt$>$0.15~GeV/c, markers are placed at the centre and not at the average of the \pt leading bin.}
    \label{fig:9}
\end{figure}

The measurement of the forward energy as a function of the leading particle \pt measured at midrapidity is complementary to the UE, defined through the charged-particle multiplicity produced in the azimuthal region transverse to the emitted leading particle ($60^{\circ}<|\Delta \varphi|<120^{\circ}$)~\cite{ALICEue0}. The charged-particle multiplicity in the transverse region rapidly reaches its maximum value and then remains constant for increasing \pt (pedestal effect), while the forward energy (both for neutrons and protons) reaches its minimum value and then remains constant for increasing particle \pt at midrapidity. These two saturation effects occur at the same leading particle \pt scale, around \pt$\sim 5$~GeV/$c$, as can be seen in Fig.~\ref{fig:9}, where UE published results~\cite{ALICEue} and ZN normalised energy are compared for pp collisions at $\sqrt{s}=13$~TeV.
In the MPI approach, collisions producing high-\pt particles have a lower than average impact parameter and consequently a larger underlying event activity~\cite{skands}. Above a leading particle \pt of about 5 GeV/$c$ the impact parameter bias reaches its maximum value.
The strong anti-correlation between the leading particle \pt\ and the forward energy seen at low $\pt^{\rm leading}$
can only be built in the initial stages of the collisions, since the two observables are causally disconnected in the following evolution stages.

\section{Conclusions}

First results on the very forward energy measured by the ALICE ZDCs in pp collisions at $\sqrt{s}=13$~TeV and in p--Pb collisions at $\sqrt{s_{\rm{NN}}}=8.16$~TeV have been presented.

In pp collisions, the energy carried by leading protons at large forward and backward rapidities is found to be uncorrelated, confirming results from hadronic collisions at lower energies, while for neutrons a weak correlation between the energy released at forward and backward rapidities is observed.

The very forward energy was studied as a function of the charged-particle production at midrapidity to gain insight in the particle production mechanisms.
The self-normalised forward energy decreases with increasing charged-particle multiplicity at midrapidity, both in pp and in p--Pb collisions with the same proton beam energy, as expected
from energy conservation.

The very forward energy decreases with increasing leading particle \pt at midrapidity until about 5~GeV/$c$ where it saturates. Similarly, the charged-particle multiplicity in the transverse region (UE) as a function of the leading particle \pt first rises and then saturates at about the same \pt value. In the case of the UE this is commonly interpreted as
a bias to an on average smaller pp impact parameter with consequently larger number of MPIs leading to higher multiplicity at midrapidity. The results of this paper
corroborate this interpretation since the correlation between central and forward rapidity can only be attributed to the initial stage of the collisions.

The hadronic interaction models used for comparison, PYTHIA 6 Perugia 2011, PYTHIA 8 Monash, and EPOS-LHC, are not able to reproduce quantitatively the measurements at large rapidities as a function of particle production at midrapidity.
These measurements provide constraints to improve the model description of beam remnants and very forward energy.

Finally, it was shown that the very forward energy can be effectively used in pp collisions as a veto to select events characterised by higher than total average multiplicity and total transverse momentum, in agreement with expectations from models including centrality dependent particle production.


\newenvironment{acknowledgement}{\relax}{\relax}
\begin{acknowledgement}
\section*{Acknowledgements}

The ALICE Collaboration would like to thank all its engineers and technicians for their invaluable contributions to the construction of the experiment and the CERN accelerator teams for the outstanding performance of the LHC complex.
The ALICE Collaboration gratefully acknowledges the resources and support provided by all Grid centres and the Worldwide LHC Computing Grid (WLCG) collaboration.
The ALICE Collaboration acknowledges the following funding agencies for their support in building and running the ALICE detector:
A. I. Alikhanyan National Science Laboratory (Yerevan Physics Institute) Foundation (ANSL), State Committee of Science and World Federation of Scientists (WFS), Armenia;
Austrian Academy of Sciences, Austrian Science Fund (FWF): [M 2467-N36] and Nationalstiftung f\"{u}r Forschung, Technologie und Entwicklung, Austria;
Ministry of Communications and High Technologies, National Nuclear Research Center, Azerbaijan;
Conselho Nacional de Desenvolvimento Cient\'{\i}fico e Tecnol\'{o}gico (CNPq), Financiadora de Estudos e Projetos (Finep), Funda\c{c}\~{a}o de Amparo \`{a} Pesquisa do Estado de S\~{a}o Paulo (FAPESP) and Universidade Federal do Rio Grande do Sul (UFRGS), Brazil;
Ministry of Education of China (MOEC) , Ministry of Science \& Technology of China (MSTC) and National Natural Science Foundation of China (NSFC), China;
Ministry of Science and Education and Croatian Science Foundation, Croatia;
Centro de Aplicaciones Tecnol\'{o}gicas y Desarrollo Nuclear (CEADEN), Cubaenerg\'{\i}a, Cuba;
Ministry of Education, Youth and Sports of the Czech Republic, Czech Republic;
The Danish Council for Independent Research | Natural Sciences, the VILLUM FONDEN and Danish National Research Foundation (DNRF), Denmark;
Helsinki Institute of Physics (HIP), Finland;
Commissariat \`{a} l'Energie Atomique (CEA) and Institut National de Physique Nucl\'{e}aire et de Physique des Particules (IN2P3) and Centre National de la Recherche Scientifique (CNRS), France;
Bundesministerium f\"{u}r Bildung und Forschung (BMBF) and GSI Helmholtzzentrum f\"{u}r Schwerionenforschung GmbH, Germany;
General Secretariat for Research and Technology, Ministry of Education, Research and Religions, Greece;
National Research, Development and Innovation Office, Hungary;
Department of Atomic Energy Government of India (DAE), Department of Science and Technology, Government of India (DST), University Grants Commission, Government of India (UGC) and Council of Scientific and Industrial Research (CSIR), India;
Indonesian Institute of Science, Indonesia;
Istituto Nazionale di Fisica Nucleare (INFN), Italy;
Institute for Innovative Science and Technology , Nagasaki Institute of Applied Science (IIST), Japanese Ministry of Education, Culture, Sports, Science and Technology (MEXT) and Japan Society for the Promotion of Science (JSPS) KAKENHI, Japan;
Consejo Nacional de Ciencia (CONACYT) y Tecnolog\'{i}a, through Fondo de Cooperaci\'{o}n Internacional en Ciencia y Tecnolog\'{i}a (FONCICYT) and Direcci\'{o}n General de Asuntos del Personal Academico (DGAPA), Mexico;
Nederlandse Organisatie voor Wetenschappelijk Onderzoek (NWO), Netherlands;
The Research Council of Norway, Norway;
Commission on Science and Technology for Sustainable Development in the South (COMSATS), Pakistan;
Pontificia Universidad Cat\'{o}lica del Per\'{u}, Peru;
Ministry of Education and Science, National Science Centre and WUT ID-UB, Poland;
Korea Institute of Science and Technology Information and National Research Foundation of Korea (NRF), Republic of Korea;
Ministry of Education and Scientific Research, Institute of Atomic Physics and Ministry of Research and Innovation and Institute of Atomic Physics, Romania;
Joint Institute for Nuclear Research (JINR), Ministry of Education and Science of the Russian Federation, National Research Centre Kurchatov Institute, Russian Science Foundation and Russian Foundation for Basic Research, Russia;
Ministry of Education, Science, Research and Sport of the Slovak Republic, Slovakia;
National Research Foundation of South Africa, South Africa;
Swedish Research Council (VR) and Knut \& Alice Wallenberg Foundation (KAW), Sweden;
European Organization for Nuclear Research, Switzerland;
Suranaree University of Technology (SUT), National Science and Technology Development Agency (NSDTA) and Office of the Higher Education Commission under NRU project of Thailand, Thailand;
Turkish Energy, Nuclear and Mineral Research Agency (TENMAK), Turkey;
National Academy of  Sciences of Ukraine, Ukraine;
Science and Technology Facilities Council (STFC), United Kingdom;
National Science Foundation of the United States of America (NSF) and United States Department of Energy, Office of Nuclear Physics (DOE NP), United States of America.
\end{acknowledgement}

\bibliographystyle{utphys}   
\bibliography{bibliography}

\newpage
\appendix

%
%

\section{The ALICE Collaboration}
\label{app:collab}

\small
\begin{flushleft}

S.~Acharya$^{\rm 143}$, 
D.~Adamov\'{a}$^{\rm 98}$, 
A.~Adler$^{\rm 76}$, 
G.~Aglieri Rinella$^{\rm 35}$, 
M.~Agnello$^{\rm 31}$, 
N.~Agrawal$^{\rm 55}$, 
Z.~Ahammed$^{\rm 143}$, 
S.~Ahmad$^{\rm 16}$, 
S.U.~Ahn$^{\rm 78}$, 
I.~Ahuja$^{\rm 39}$
% ...

\section*{Affiliation notes} 

$^{\rm I}$ Deceased\\
$^{\rm II}$ Also at: Italian National Agency for New Technologies, Energy and Sustainable Economic Development (ENEA), Bologna, Italy\\
$^{\rm III}$ Also at: Dipartimento DET del Politecnico di Torino, Turin, Italy\\
$^{\rm IV}$ Also at: M.V. Lomonosov Moscow State University, D.V. Skobeltsyn Institute of Nuclear, Physics, Moscow, Russia\\
$^{\rm V}$ Also at: Department of Applied Physics, Aligarh Muslim University, Aligarh, India\\
$^{\rm VI}$ Also at: Institute of Theoretical Physics, University of Wroclaw, Poland\\

\section*{Collaboration Institutes}

$^{1}$ A.I. Alikhanyan National Science Laboratory (Yerevan Physics Institute) Foundation, Yerevan, Armenia\\
$^{2}$ AGH University of Science and Technology, Cracow, Poland\\
$^{3}$ Bogolyubov Institute for Theoretical Physics, National Academy of Sciences of Ukraine, Kiev, Ukraine\\
$^{4}$ Bose Institute, Department of Physics  and Centre for Astroparticle Physics and Space Science (CAPSS), Kolkata, India\\
$^{5}$ Budker Institute for Nuclear Physics, Novosibirsk, Russia\\
$^{6}$ California Polytechnic State University, San Luis Obispo, California, United States\\
$^{7}$ Central China Normal University, Wuhan, China\\
$^{8}$ Centro de Aplicaciones Tecnol\'{o}gicas y Desarrollo Nuclear (CEADEN), Havana, Cuba\\
$^{9}$ Centro de Investigaci\'{o}n y de Estudios Avanzados (CINVESTAV), Mexico City and M\'{e}rida, Mexico\\
$^{10}$ Chicago State University, Chicago, Illinois, United States\\
$^{11}$ China Institute of Atomic Energy, Beijing, China\\
$^{12}$ Chungbuk National University, Cheongju, Republic of Korea\\
$^{13}$ Comenius University Bratislava, Faculty of Mathematics, Physics and Informatics, Bratislava, Slovakia\\
$^{14}$ COMSATS University Islamabad, Islamabad, Pakistan\\
$^{15}$ Creighton University, Omaha, Nebraska, United States\\
$^{16}$ Department of Physics, Aligarh Muslim University, Aligarh, India\\
$^{17}$ Department of Physics, Pusan National University, Pusan, Republic of Korea\\
$^{18}$ Department of Physics, Sejong University, Seoul, Republic of Korea\\
$^{19}$ Department of Physics, University of California, Berkeley, California, United States\\
$^{20}$ Department of Physics, University of Oslo, Oslo, Norway\\
$^{21}$ Department of Physics and Technology, University of Bergen, Bergen, Norway\\
$^{22}$ Dipartimento di Fisica dell'Universit\`{a} 'La Sapienza' and Sezione INFN, Rome, Italy\\
$^{23}$ Dipartimento di Fisica dell'Universit\`{a} and Sezione INFN, Cagliari, Italy\\
$^{24}$ Dipartimento di Fisica dell'Universit\`{a} and Sezione INFN, Trieste, Italy\\
$^{25}$ Dipartimento di Fisica dell'Universit\`{a} and Sezione INFN, Turin, Italy\\
$^{26}$ Dipartimento di Fisica e Astronomia dell'Universit\`{a} and Sezione INFN, Bologna, Italy\\
$^{27}$ Dipartimento di Fisica e Astronomia dell'Universit\`{a} and Sezione INFN, Catania, Italy\\
$^{28}$ Dipartimento di Fisica e Astronomia dell'Universit\`{a} and Sezione INFN, Padova, Italy\\
$^{29}$ Dipartimento di Fisica e Nucleare e Teorica, Universit\`{a} di Pavia, Pavia, Italy\\
$^{30}$ Dipartimento di Fisica `E.R.~Caianiello' dell'Universit\`{a} and Gruppo Collegato INFN, Salerno, Italy\\
$^{31}$ Dipartimento DISAT del Politecnico and Sezione INFN, Turin, Italy\\
$^{32}$ Dipartimento di Scienze e Innovazione Tecnologica dell'Universit\`{a} del Piemonte Orientale and INFN Sezione di Torino, Alessandria, Italy\\
$^{33}$ Dipartimento di Scienze MIFT, Universit\`{a} di Messina, Messina, Italy\\
$^{34}$ Dipartimento Interateneo di Fisica `M.~Merlin' and Sezione INFN, Bari, Italy\\
$^{35}$ European Organization for Nuclear Research (CERN), Geneva, Switzerland\\
$^{36}$ Faculty of Electrical Engineering, Mechanical Engineering and Naval Architecture, University of Split, Split, Croatia\\
$^{37}$ Faculty of Engineering and Science, Western Norway University of Applied Sciences, Bergen, Norway\\
$^{38}$ Faculty of Nuclear Sciences and Physical Engineering, Czech Technical University in Prague, Prague, Czech Republic\\
$^{39}$ Faculty of Science, P.J.~\v{S}af\'{a}rik University, Ko\v{s}ice, Slovakia\\
$^{40}$ Frankfurt Institute for Advanced Studies, Johann Wolfgang Goethe-Universit\"{a}t Frankfurt, Frankfurt, Germany\\
$^{41}$ Fudan University, Shanghai, China\\
$^{42}$ Gangneung-Wonju National University, Gangneung, Republic of Korea\\
$^{43}$ Gauhati University, Department of Physics, Guwahati, India\\
$^{44}$ Helmholtz-Institut f\"{u}r Strahlen- und Kernphysik, Rheinische Friedrich-Wilhelms-Universit\"{a}t Bonn, Bonn, Germany\\
$^{45}$ Helsinki Institute of Physics (HIP), Helsinki, Finland\\
$^{46}$ High Energy Physics Group,  Universidad Aut\'{o}noma de Puebla, Puebla, Mexico\\
$^{47}$ Hiroshima University, Hiroshima, Japan\\
$^{48}$ Hochschule Worms, Zentrum  f\"{u}r Technologietransfer und Telekommunikation (ZTT), Worms, Germany\\
$^{49}$ Horia Hulubei National Institute of Physics and Nuclear Engineering, Bucharest, Romania\\
$^{50}$ Indian Institute of Technology Bombay (IIT), Mumbai, India\\
$^{51}$ Indian Institute of Technology Indore, Indore, India\\
$^{52}$ Indonesian Institute of Sciences, Jakarta, Indonesia\\
$^{53}$ INFN, Laboratori Nazionali di Frascati, Frascati, Italy\\
$^{54}$ INFN, Sezione di Bari, Bari, Italy\\
$^{55}$ INFN, Sezione di Bologna, Bologna, Italy\\
$^{56}$ INFN, Sezione di Cagliari, Cagliari, Italy\\
$^{57}$ INFN, Sezione di Catania, Catania, Italy\\
$^{58}$ INFN, Sezione di Padova, Padova, Italy\\
$^{59}$ INFN, Sezione di Pavia, Pavia, Italy\\
$^{60}$ INFN, Sezione di Roma, Rome, Italy\\
$^{61}$ INFN, Sezione di Torino, Turin, Italy\\
$^{62}$ INFN, Sezione di Trieste, Trieste, Italy\\
$^{63}$ Inha University, Incheon, Republic of Korea\\
$^{64}$ Institute for Gravitational and Subatomic Physics (GRASP), Utrecht University/Nikhef, Utrecht, Netherlands\\
$^{65}$ Institute for Nuclear Research, Academy of Sciences, Moscow, Russia\\
$^{66}$ Institute of Experimental Physics, Slovak Academy of Sciences, Ko\v{s}ice, Slovakia\\
$^{67}$ Institute of Physics, Homi Bhabha National Institute, Bhubaneswar, India\\
$^{68}$ Institute of Physics of the Czech Academy of Sciences, Prague, Czech Republic\\
$^{69}$ Institute of Space Science (ISS), Bucharest, Romania\\
$^{70}$ Institut f\"{u}r Kernphysik, Johann Wolfgang Goethe-Universit\"{a}t Frankfurt, Frankfurt, Germany\\
$^{71}$ Instituto de Ciencias Nucleares, Universidad Nacional Aut\'{o}noma de M\'{e}xico, Mexico City, Mexico\\
$^{72}$ Instituto de F\'{i}sica, Universidade Federal do Rio Grande do Sul (UFRGS), Porto Alegre, Brazil\\
$^{73}$ Instituto de F\'{\i}sica, Universidad Nacional Aut\'{o}noma de M\'{e}xico, Mexico City, Mexico\\
$^{74}$ iThemba LABS, National Research Foundation, Somerset West, South Africa\\
$^{75}$ Jeonbuk National University, Jeonju, Republic of Korea\\
$^{76}$ Johann-Wolfgang-Goethe Universit\"{a}t Frankfurt Institut f\"{u}r Informatik, Fachbereich Informatik und Mathematik, Frankfurt, Germany\\
$^{77}$ Joint Institute for Nuclear Research (JINR), Dubna, Russia\\
$^{78}$ Korea Institute of Science and Technology Information, Daejeon, Republic of Korea\\
$^{79}$ KTO Karatay University, Konya, Turkey\\
$^{80}$ Laboratoire de Physique des 2 Infinis, Ir\`{e}ne Joliot-Curie, Orsay, France\\
$^{81}$ Laboratoire de Physique Subatomique et de Cosmologie, Universit\'{e} Grenoble-Alpes, CNRS-IN2P3, Grenoble, France\\
$^{82}$ Lawrence Berkeley National Laboratory, Berkeley, California, United States\\
$^{83}$ Lund University Department of Physics, Division of Particle Physics, Lund, Sweden\\
$^{84}$ Moscow Institute for Physics and Technology, Moscow, Russia\\
$^{85}$ Nagasaki Institute of Applied Science, Nagasaki, Japan\\
$^{86}$ Nara Women{'}s University (NWU), Nara, Japan\\
$^{87}$ National and Kapodistrian University of Athens, School of Science, Department of Physics , Athens, Greece\\
$^{88}$ National Centre for Nuclear Research, Warsaw, Poland\\
$^{89}$ National Institute of Science Education and Research, Homi Bhabha National Institute, Jatni, India\\
$^{90}$ National Nuclear Research Center, Baku, Azerbaijan\\
$^{91}$ National Research Centre Kurchatov Institute, Moscow, Russia\\
$^{92}$ Niels Bohr Institute, University of Copenhagen, Copenhagen, Denmark\\
$^{93}$ Nikhef, National institute for subatomic physics, Amsterdam, Netherlands\\
$^{94}$ NRC Kurchatov Institute IHEP, Protvino, Russia\\
$^{95}$ NRC \guillemotleft Kurchatov\guillemotright  Institute - ITEP, Moscow, Russia\\
$^{96}$ NRNU Moscow Engineering Physics Institute, Moscow, Russia\\
$^{97}$ Nuclear Physics Group, STFC Daresbury Laboratory, Daresbury, United Kingdom\\
$^{98}$ Nuclear Physics Institute of the Czech Academy of Sciences, \v{R}e\v{z} u Prahy, Czech Republic\\
$^{99}$ Oak Ridge National Laboratory, Oak Ridge, Tennessee, United States\\
$^{100}$ Ohio State University, Columbus, Ohio, United States\\
$^{101}$ Petersburg Nuclear Physics Institute, Gatchina, Russia\\
$^{102}$ Physics department, Faculty of science, University of Zagreb, Zagreb, Croatia\\
$^{103}$ Physics Department, Panjab University, Chandigarh, India\\
$^{104}$ Physics Department, University of Jammu, Jammu, India\\
$^{105}$ Physics Department, University of Rajasthan, Jaipur, India\\
$^{106}$ Physikalisches Institut, Eberhard-Karls-Universit\"{a}t T\"{u}bingen, T\"{u}bingen, Germany\\
$^{107}$ Physikalisches Institut, Ruprecht-Karls-Universit\"{a}t Heidelberg, Heidelberg, Germany\\
$^{108}$ Physik Department, Technische Universit\"{a}t M\"{u}nchen, Munich, Germany\\
$^{109}$ Politecnico di Bari and Sezione INFN, Bari, Italy\\
$^{110}$ Research Division and ExtreMe Matter Institute EMMI, GSI Helmholtzzentrum f\"ur Schwerionenforschung GmbH, Darmstadt, Germany\\
$^{111}$ Russian Federal Nuclear Center (VNIIEF), Sarov, Russia\\
$^{112}$ Saha Institute of Nuclear Physics, Homi Bhabha National Institute, Kolkata, India\\
$^{113}$ School of Physics and Astronomy, University of Birmingham, Birmingham, United Kingdom\\
$^{114}$ Secci\'{o}n F\'{\i}sica, Departamento de Ciencias, Pontificia Universidad Cat\'{o}lica del Per\'{u}, Lima, Peru\\
$^{115}$ St. Petersburg State University, St. Petersburg, Russia\\
$^{116}$ Stefan Meyer Institut f\"{u}r Subatomare Physik (SMI), Vienna, Austria\\
$^{117}$ SUBATECH, IMT Atlantique, Universit\'{e} de Nantes, CNRS-IN2P3, Nantes, France\\
$^{118}$ Suranaree University of Technology, Nakhon Ratchasima, Thailand\\
$^{119}$ Technical University of Ko\v{s}ice, Ko\v{s}ice, Slovakia\\
$^{120}$ The Henryk Niewodniczanski Institute of Nuclear Physics, Polish Academy of Sciences, Cracow, Poland\\
$^{121}$ The University of Texas at Austin, Austin, Texas, United States\\
$^{122}$ Universidad Aut\'{o}noma de Sinaloa, Culiac\'{a}n, Mexico\\
$^{123}$ Universidade de S\~{a}o Paulo (USP), S\~{a}o Paulo, Brazil\\
$^{124}$ Universidade Estadual de Campinas (UNICAMP), Campinas, Brazil\\
$^{125}$ Universidade Federal do ABC, Santo Andre, Brazil\\
$^{126}$ University of Cape Town, Cape Town, South Africa\\
$^{127}$ University of Houston, Houston, Texas, United States\\
$^{128}$ University of Jyv\"{a}skyl\"{a}, Jyv\"{a}skyl\"{a}, Finland\\
$^{129}$ University of Kansas, Lawrence, Kansas, United States\\
$^{130}$ University of Liverpool, Liverpool, United Kingdom\\
$^{131}$ University of Science and Technology of China, Hefei, China\\
$^{132}$ University of South-Eastern Norway, Tonsberg, Norway\\
$^{133}$ University of Tennessee, Knoxville, Tennessee, United States\\
$^{134}$ University of the Witwatersrand, Johannesburg, South Africa\\
$^{135}$ University of Tokyo, Tokyo, Japan\\
$^{136}$ University of Tsukuba, Tsukuba, Japan\\
$^{137}$ Universit\'{e} Clermont Auvergne, CNRS/IN2P3, LPC, Clermont-Ferrand, France\\
$^{138}$ Universit\'{e} de Lyon, CNRS/IN2P3, Institut de Physique des 2 Infinis de Lyon , Lyon, France\\
$^{139}$ Universit\'{e} de Strasbourg, CNRS, IPHC UMR 7178, F-67000 Strasbourg, France, Strasbourg, France\\
$^{140}$ Universit\'{e} Paris-Saclay Centre d'Etudes de Saclay (CEA), IRFU, D\'{e}partment de Physique Nucl\'{e}aire (DPhN), Saclay, France\\
$^{141}$ Universit\`{a} degli Studi di Foggia, Foggia, Italy\\
$^{142}$ Universit\`{a} di Brescia, Brescia, Italy\\
$^{143}$ Variable Energy Cyclotron Centre, Homi Bhabha National Institute, Kolkata, India\\
$^{144}$ Warsaw University of Technology, Warsaw, Poland\\
$^{145}$ Wayne State University, Detroit, Michigan, United States\\
$^{146}$ Westf\"{a}lische Wilhelms-Universit\"{a}t M\"{u}nster, Institut f\"{u}r Kernphysik, M\"{u}nster, Germany\\
$^{147}$ Wigner Research Centre for Physics, Budapest, Hungary\\
$^{148}$ Yale University, New Haven, Connecticut, United States\\
$^{149}$ Yonsei University, Seoul, Republic of Korea\\

\end{flushleft} 

\end{document}